\documentclass[12pt]{article}
\usepackage{amsmath}
\usepackage{graphicx}
\usepackage{enumerate}
\usepackage{natbib}
\usepackage{url} 
\usepackage[colorlinks=true,allcolors=blue]{hyperref}
\usepackage{url}
\usepackage{graphicx}
\usepackage{amsmath, amssymb}
\usepackage{bbold}
\usepackage{mathbbol}
\usepackage{algorithm}
\usepackage{algorithmic}
\usepackage{enumitem}
\usepackage{float}
\usepackage{xcolor}
\usepackage{bm} 
\usepackage{setspace}
\usepackage{multirow}
\usepackage{lscape}
\usepackage{rotating}
\usepackage{adjustbox}
\usepackage{booktabs}
\usepackage{ragged2e}

\usepackage{caption}
\usepackage{subcaption}
\usepackage{float}
\def\*#1{\mathbf{#1}}
\def\+#1{\amsmathbb{#1}}
\def\^#1{\mathbb{#1}}
\DeclareSymbolFontAlphabet{\amsmathbb}{AMSb}%

\newcommand{\blind}{1}

\begin{document}

\def\spacingset#1{\renewcommand{\baselinestretch}%
{#1}\small\normalsize} \spacingset{1}


\if1\blind
{
  \title{\bf  Functional proportional hazards mixture cure model and its application to modelling the association between cancer mortality and physical activity in NHANES 2003-2006}
  \author{Rahul Ghosal$^1$, Marcos Matabuena$^{2}$, Jiajia Zhang$^{1}$ \hspace{.2cm}\\
     $^{1}$ Department of Epidemiology and Biostatistics, University of South Carolina \\
$^{2}$ Centro Singular de Investigación en Tecnologías Intelixentes, \\Universidad de Santiago de Compostela, Santiago de Compostela, Spain\\
}
  \maketitle
} \fi

\if0\blind
{
  \bigskip
  \bigskip
  \bigskip
  \begin{center}
    {\LARGE\bf Functional proportinal hazards mixture cure model}
\end{center}
  \medskip
} \fi

\bigskip
\begin{abstract}
We develop a functional proportional hazards mixture cure (FPHMC) model with scalar and functional covariates measured at the baseline. The mixture cure model, useful in studying populations with a cure fraction of a particular event of interest is extended to functional data. We employ the EM algorithm and develop a semiparametric penalized spline-based approach to estimate the dynamic functional coefficients of the incidence and the latency part. The proposed method is computationally efficient and simultaneously incorporates smoothness in the estimated functional coefficients via roughness penalty. Simulation studies illustrate a satisfactory performance of the proposed method in accurately estimating the model parameters and the baseline survival function. 
Finally, the clinical potential of the model is demonstrated in two real data examples that incorporate rich high-dimensional biomedical signals as functional covariates measured at the baseline and constitute novel domains to apply cure survival models in contemporary medical situations. In particular, we analyze i)  minute-by-minute physical activity data from the National Health and Nutrition Examination Survey (NHANES) 2003-2006 to study the association between diurnal patterns of physical activity (PA) at baseline and all cancer mortality through 2019 while adjusting for other biological factors; ii) the impact of daily functional measures of disease severity collected in the intensive care unit on post ICU recovery and mortality event. Our findings provide novel epidemiological insights into the association between daily patterns of PA and cancer mortality.
Software implementation and illustration of the proposed estimation method is provided in R.

\end{abstract}

\noindent%
\vfill

\newpage
\spacingset{1.9} 
\section{Introduction}
\label{sec:intro4}

Technological improvements in wearable devices allow us to capture real-time, continuous and high-resolution streams of user-specific physiological data such as minute-level step counts, heart rate (beats per minute and ECG), energy expenditure (EE), brainwave (EEG), and many others. This rich source of information provides new opportunities in the predictive models for a deeper understanding of human behaviors and their influence on human health and disease. 
Functional data analysis \citep{Ramsay05functionaldata} is a branch of statistics that analyzes data over a continuum. It allows us to answer clinical questions more efficiently about continuous biomedical signals and their impact on human health.
Functional data analysis has seen diverse applications in biological sciences such as genome-wide association studies (GWAS) \citep{fan2017high}, physical activity research \citep{goldsmith2016new,ghosal2021variable,cui2022fast,matabuena2022estimating}, functional magnetic resonance imaging \citep{reiss2017methods} and many others. In addition, several regression methods exist in functional data analysis (FDA) literature to model scalar and functional outcomes of interest \citep{huang2004polynomial, reiss2010fast,reiss2017methods}.

In many biological applications, practitioners are often more concerned with survival or time-to-event outcomes and their association with risk factors of interest \citep{https://doi.org/10.48550/arxiv.2206.06885}. There exists a rich literature in survival analysis with parametric \citep{cai2009regularized}, semi-parametric \citep{cheng1997predicting} and nonparametric models \citep{dabrowska1987non} for modeling survival outcomes based on a set of scalar predictors, such as the proportional hazards (PH) model \citep{cox1972regression} and the proportional odds (PO) model \citep{bennett1983analysis}. Cure survival models are a relatively recent approach in the literature that assumes that there exists a cure fraction in the population, subjects who will not experience the event of interest in the future. For example, this situation might arise in many oncological studies. In such cases, the systematic use of traditional survival models such as PH or PO is unrealistic as they cannot accommodate the number of cured patients. In the presence of cure fraction, the two-component mixture cure model \citep{farewell1982use} and the promotion time cure model \citep{tsodikov2003estimating} are the primarily used approaches to model survival times. The two-component mixture cure model
is popular due to its interpretability and models the
cured and uncured proportion directly by the incidence and latency
part. Depending on the survival models used in the latency part, there have been many proposals in the mixture cure model literature such as the proportional hazards mixture cure (PHMC) model \citep{kuk1992mixture,peng2000nonparametric,sy2000estimation}, proportional odds
mixture cure (POMC) model \citep{gu2011analysis} and generalized odds rate mixture cure (GORMC)
model \citep{zhou2018computationally}.  Non-parametric approaches give another important research direction \citep{lopez2017nonparametric} based on the Beran estimator for the survival component and the Nadaraya-Watson estimator for the cure. Still, their application in epidemiological studies can be limited due to the need for availability of large sample sizes, especially when the dimension of the predictor is large.

\begin{figure}[ht]
	\centering
 \begin{subfigure}[b]{0.48\textwidth}
         \centering
        \includegraphics[width=0.99\textwidth,height=1\textwidth]{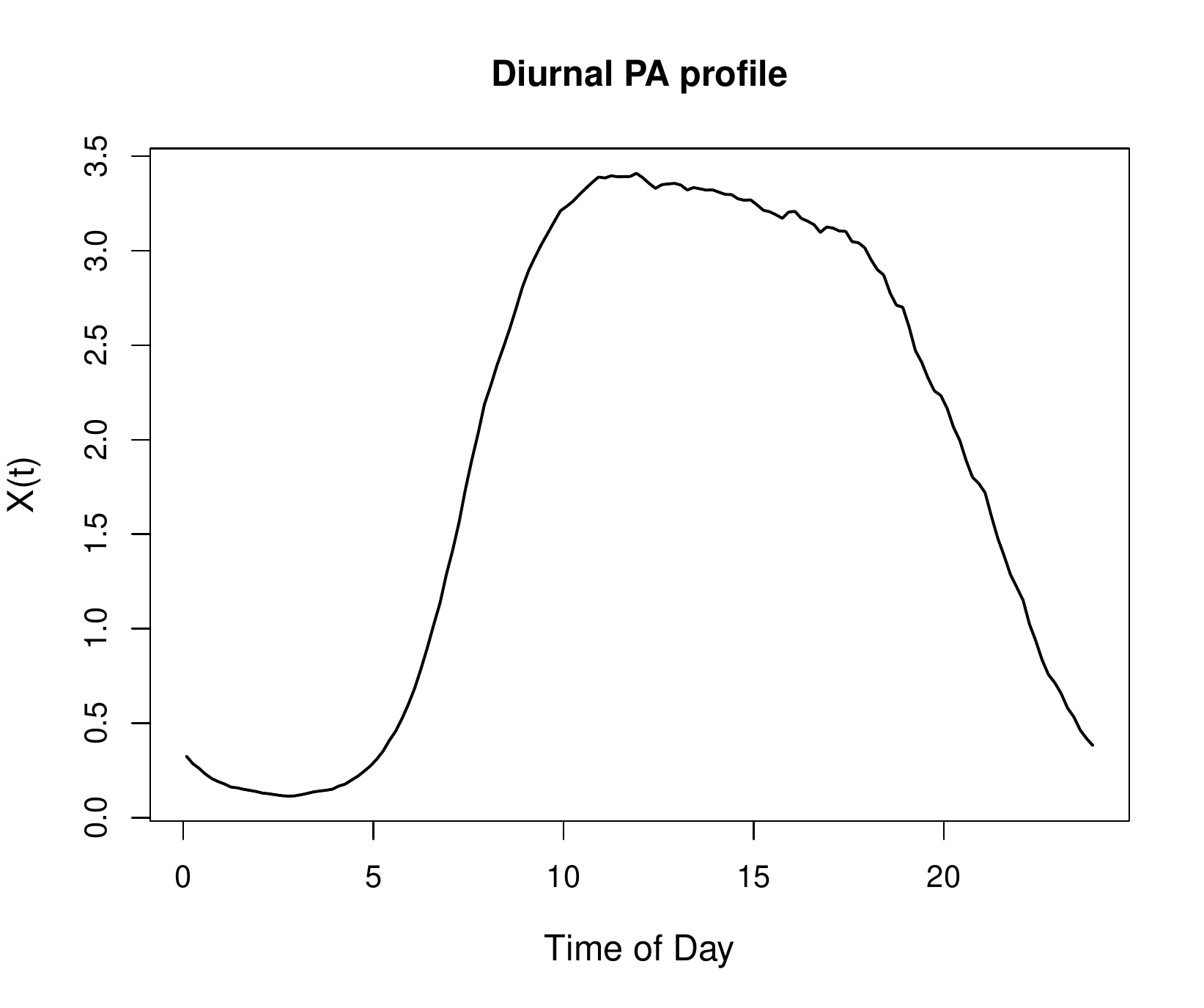}
\caption{Average diurnal physical activity profile for NHANES participants.} 
\label{fig:grafactividad1}	
     \end{subfigure}
     \hfill
     \begin{subfigure}[b]{0.5\textwidth}
        \centering\includegraphics[width=0.99\textwidth]{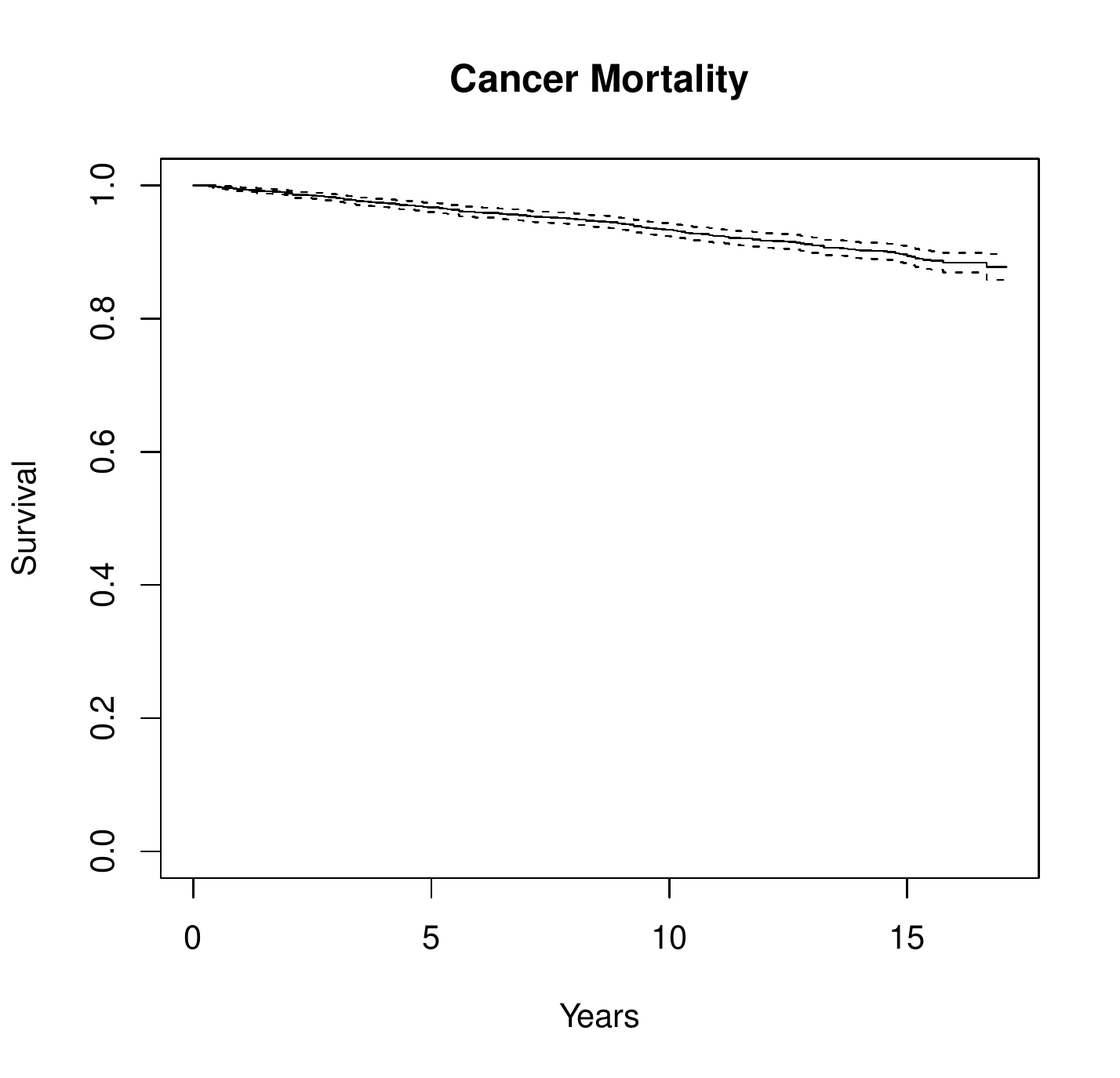}
\caption{Kaplan-Meier marginal survival curve of all cancer mortality in the NHANES analysis.}
         \label{fig:kmage1}
     \end{subfigure}
     \caption{Diurnal PA profile and Kaplan Meier survival curves in the NHANES analysis.}
\end{figure}

Although we are increasingly collecting high dimensional functional observations, e.g., physical activity (PA), heart-rate (HR), EE, and many such biomedical signals, literature on functional data analysis with survival outcomes and functional covariates have been relatively sparser. In this paper, as an motivating application we consider studying the association between diurnal patterns of objectively measured physical activity data collected from elderly participants (aged 50-85 years) in the National Health and Nutrition Examination Survey (NHANES) 2003-2006 at baseline, and all-cancer mortality up untill 2019 while adjusting for other biological factors. Previous research exploring relation between PA and cancer have primarily explored various summary measures of PA \citep{liu2016leisure,patel2019american}. Figure \ref{fig:grafactividad1} displays the average diurnal PA profile across all participants for log-transformed activity counts. 
Analysis of the marginal Kaplan-Meier curve (Figure \ref{fig:kmage1}), indicates the presence of a cure fraction. From a biological point of view, this can be explained as i) there exists a fraction of subjects who might be at a higher risk of cancer mortality in the time-period considered due to age, sedentary lifestyle, and other biological factors, while ii) a large fraction of the subjects, will not experience the event of interest i.e., death due to cancer. 

The majority of the focus in survival models with functional data  has been on the linear functional Cox model \citep{gellar2015cox,qu2016optimal,kong2018flcrm}, which can capture a linear functional effect of the functional covariate on the log-hazard function. \cite{cui2021additive} extended these ideas to an additive functional cox model for quantifying the additive and nonlinear association between a time-to-event outcome and functional covariates. The first functional quantile survival model has been proposed in \cite{quantilefunctional}. However, as mentioned earlier, these methods are no longer applicable in the presence of a cure fraction in the population. In this article, we develop a new mixture cure model for survival outcomes involving multiple scalars and one or more functional covariates in the presence of right censoring. Under the proportional hazards assumption, we propose a functional proportional hazards mixture cure (FPHMC) model with scalar and functional covariates measured at the baseline. To the best of our knowledge in the rich literature on cure models, there is only one work \citep{shi2022} in this direction for interval-censored data based on a two-stage functional principal component analysis (FPCA) based approach, which does not involve a smoothing process in the model formulation. We employ the EM algorithm and develop a semiparametric penalized spline-based approach to estimate the functional coefficients of the incidence and the latency part. The incidence model is shown to reduce to a scalar-on functional regression model \citep{reiss2017methods}, and the latency model resembles a linear functional cox model. The proposed method is computationally efficient, and use of roughness penalty simultaneously incorporates smoothness in the estimated functional coefficients of the incidence and latency part. 



The key contributions of this paper include i) Development of a functional proportional hazard mixture cure model with multiple scalar and one or more functional covariates  ii) Incorporation of smoothness in the estimated functional coefficients of the latency and incidence part with automated smoothing parameter selection, which avoids selection of number of principal components in an FPCA based approach \citep{shi2022}. 
iii) A novel application on NHANES data capturing the effect of daily pattern of PA on cancer mortality, after adjusting for age and other biological confounders.
Numerical analyses using simulations illustrate the proposed method's satisfactory finite-sample performance in terms of estimation accuracy of the scalar and functional coefficients and the baseline survival function compared to competing methods. The applications on  real data  provide novel epidemiological insights into the association between the functional, scalar biomarkers and the latency as well as the incidence of the time-to-event outcomes.

The rest of this article is organized in the following way. The modeling framework of the functional proportional hazards mixture
cure (FPHMC) model is presented in Section 2, along with the proposed EM-based estimation method. Section 3 presents simulation studies and illustrates the finite sample performance of the proposed method. Real data applications of the proposed method are demonstrated in Section 4. We deliberate on the contributions of the proposed method  and some possible extensions of this work in Section 5.

\section{Methodology}
Let $T_i$ be the survival time for subject i, and $C_i$ the corresponding censoring time for $i=1,2,\ldots,n$. We assume independent right censoring and denote $Y_i= min(T_i,C_i)$ (observed survival time for subject $i$), $\Delta_i= I(T_i\leq C_i)$ (the event or censoring indicator for subject $i$). Suppose that we observe a random sample $(Y_i,\Delta_i, \*X_i, \*Z_i, X_{Si}(\cdot),Z_{Si}(\cdot))$, $i=1,\dots, n$, having the same distribution as $(Y,\Delta, \*X,\*Z, X_{S}(\cdot),Z_{S}(\cdot))$. Functional covariates $X_{Si}(\cdot),Z_{Si}(\cdot) $ are observed at the baseline and assumed to lie in a real separable Hilbert space, which is taken to be $L^2[0,1]$ in this paper and $\*X_i,\*Z_i$ are usual scalar covariates. 
Let us assume both the functional covariates $X_{Si}(\cdot),Z_{Si}(\cdot)$ are observed on $\mathcal{S}=[0,1]$ for simplicity, although the presented model and estimation method allow for different domains.
In developing our method, we assume that the functional covariates are observed on a dense and regular grid $S= \{s_{1},s_{2},\ldots,s_{m} \} \subset \mathcal{S}=[0,1] $, although this can be easily extended to accommodate more general scenarios, e.g., covariates observed on irregular and sparse domain.
We assume the censoring time $C_i$ and event time $T_i$ are independent, conditional on the other covariates.

We posit the following mixture cure model for the overall survival function of subject $i$.
\begin{equation}
    S(t|x_i,z_i,x_{si}(\cdot),z_{si}(\cdot))= 1-\pi(\*z_i,z_{si}(\cdot))+\pi(\*z_i,z_{si}(\cdot))S_{u}(t|\*x_i,x_{si}(\cdot)).
\end{equation}
Here $\pi(\*z_i,z_{si}(\cdot))= P(B_i=1|\*Z_i=\*z_i,Z_{Si}(\cdot)=z_{si}(\cdot))$ is the probability of being susceptible (often called the incidence of the model), and $S_{u}(t|\*x_i,x_{si}(\cdot))= P(T_i>t|\*X_i=\*x_i,X_{Si}(\cdot)=x_{si}(\cdot),B_i=1)$ is the conditional survival function of susceptibles (often called the latency of the model). Here, $B_i= I(T_i<\infty)$ is the latent binary variable indicating whether subject $i$ is susceptible or cured. Like the usual mixture cure model, the incidence and latency part can be modelled separately. The proposed modelling framework allows $\*X_i,\*Z_i$ and $ X_{Si}(\cdot),Z_{Si}(\cdot))$ to be the same covariates as in usual mixture cure model model, moreover it contains the cases where there might not be any functional covariates in the incidence or latency part (or both). Finally, we assume that $X_{Si}(\cdot),Z_{Si}(\cdot)$ are centered without loss of generality. 

\subsection*{Cure Submodel}
In the cure submodel, we assume the latent binary variables $B_1,B_2,\ldots,B_n$ comes from an exponential family. We consider the following generalized scalar-on-function regression 
\begin{eqnarray}
E(B_i|\*Z_i=\*z_i,Z_{Si}(\cdot)=z_{si}(\cdot))=\mu_i=\pi(\*z_i,z_{si}(\cdot)), \hspace{2 mm}
g(\mu_i)=\*z_i^T\*b + \int_{0}^{1} z_{si}(s)b(s)ds.\label{cure1}
\end{eqnarray}
Here $g$ is a known link function, e.g., the logit link $g(p)=log(\frac{p}{1-p})$, the probit link $g(p)=\Phi^{-1}(p)$ ($\Phi$ is the standard Normal c.d.f) or the complementary log-log link 
$g(p)=log(-log(1-p))$. The intercept term is included within $\*z_i$ and $\*b$ captures the effect of the baseline scalar covariates on incidence. The functional-coefficient $b(s)$ captures the dynamic effect of the baseline functional covariate $z_s(\cdot)$ on incidence.

\subsection*{Latency Submodel}
For the latency part, we make the proportional hazards (PH) assumption, and consider the following linear functional Cox model
\begin{equation}
    log \lambda_{u}(t|\*x_i,x_{si}(\cdot))=log \lambda_{0}(t)+\*x_i^T\bm\beta + \int_{0}^{1}x_{si}(s)\beta(s)ds.\label{lat1}
\end{equation}
Here $\lambda_u(\cdot|\*x_i,x_{si}(\cdot))$ denotes the hazard function of the susceptibles corresponding to $S_u(\cdot|\*x_i,x_{si}(\cdot))$. Let $S_0(t)$ denote the baseline survival function of susceptibles given by $S_0(t)=P(T>t|\*X=\*0,X_{S}(\cdot)=0,B=1)$. Then $\lambda_{0}(t)$ is the corresponding baseline hazard function and $S_0(t)=exp(-\int_{0}^{t}\lambda_0(u)du )$. The parameter $\bm\beta$ captures the effect of the baseline scalar covariates on the survival hazard of the susceptibles and the functional coefficient $\beta(s)$ captures the dynamic effect of the baseline covariate $x_s(\cdot)$ on the hazard of the susceptibles. In particular, $exp(\int_{0}^{1}\beta(s)ds)$ is the multiplicative increase in the hazard of susceptibles if the entire covariate $x_s(\cdot)$ is increased by one unit \citep{gellar2015cox}.

\subsection{Estimation}
Let us denote the unknown parameters by $\bm\Theta=(\*b,b(\cdot),\bm\beta,\beta(\cdot),S_0(t))$, where $S_0(t)$ is the baseline survival function and define $\*O=(\*O_1,\*O_2,\ldots,\*O_n)$,  $\*B=(B_1,B_2,\ldots,B_n)$. 
Since the latent binary variables $B_i$s are unobserved, we follow an EM based estimation approach \citep{peng2000nonparametric,sy2000estimation} for estimation of the parameters of interest in the FPHMC model. Let us denote the observed data for $i$-th subject as $\*O_i=(Y_i,\Delta_i, \*x_i, \*z_i, x_{si}(\cdot),z_{si}(\cdot))$. 
The complete likelihood function is given by,
\begin{equation}
    L_c(\bm\Theta;\*O,\*B)=\prod_{i=1}^{n} (1-\pi(\*z_i,z_{si}(\cdot)))^{1-B_i}\pi(\*z_i,z_{si}(\cdot))^{B_i}\lambda_u(Y_i|\*x_i,x_{si}(\cdot))^{\Delta_iB_i}S_u(Y_i|\*x_i,x_{si}(\cdot))^{B_i}.
\end{equation}
The logarithm of the complete likelihood separates into sum of two factors,
\begin{eqnarray}
      l_c(\bm\Theta;\*O,\*B)= l_{c_1}(\*b,b(\cdot);\*O,\*B)+ l_{c_2}(\bm\beta,\beta(\cdot),S_0(t);\*O,\*B)\\
    l_{c_1}(\*b,b(\cdot);\*O,\*B)=\sum_{i=1}^{n}B_ilog\{\pi(\*z_i,z_{si}(\cdot))\}+(1-B_i)log\{1-\pi(\*z_i,z_{si}(\cdot))\} \label{part1}\\ 
      l_{c_2}(\bm\beta,\beta(\cdot),S_0(t);\*O,\*B)=\sum_{i=1}^{n}\Delta_iB_ilog\{\lambda_u(Y_i|\*x_i,x_{si}(\cdot))\}+B_ilog\{S_u(Y_i|\*x_i,x_{si}(\cdot))\}.\label{part2}
\end{eqnarray}
We use the following EM algorithm to estimate the parameters $\bm\Theta$.

\subsection*{E-Step}
In the E-Step of the EM algorithm, we calculate the conditional expectation of the complete loglikelihood with respect to $B_i$ given the observed data $\*O$ and current parameter estimates $\bm\Theta^{(m)}=(\*b^{(m)},b^{(m)}(\cdot),\bm\beta^{(m)},\beta^{(m)}(\cdot),S_0^{(m)}(t))$. As both (\ref{part1}) and (\ref{part2}) are linear functions of $B_i$, it is enough to compute the conditional expectation $E(B_i|\*O,\bm\Theta^{(m)})$. In particular, we have,  
\begin{equation}
E(B_i|\*O,\bm\Theta^{(m)})=w_i^{(m)}=\Delta_i+(1-\Delta_i)\frac{\pi(\*z_i,z_{si}(\cdot))S_{u}(Y_i|\*x_i,x_{si}(\cdot))}{1-\pi(\*z_i,z_{si}(\cdot))  +\pi(\*z_i,z_{si}(\cdot))S_{u}(Y_i|\*x_i,x_{si}(\cdot))}.
\end{equation}
Note that $w_i^{(m)}=1$ if $\Delta_i=1$, hence $\Delta_ilogw_i^{(m)}=0$ and $\Delta_iw_i^{(m)}=\Delta_i$. The expected complete loglikelihood is obtained by plugging in these expressions in (\ref{part1}) and (\ref{part2}). The two expectations become,
\begin{eqnarray}
    E(l_{c_1}(\*b,b(\cdot);\*O,\*B))=\sum_{i=1}^{n}w_i^{(m)}log\{\pi(\*z_i,z_{si}(\cdot))\}+(1-w_i^{(m)})log\{1-\pi(\*z_i,z_{si}(\cdot))\} \label{E1}\\
    E(l_{c_2}(\bm\beta,\beta(\cdot),S_0(t);\*O,\*B))=
    \sum_{i=1}^{n}\Delta_ilog\{w_i^{(m)}\lambda_u(Y_i|\*x_i,x_{si}(\cdot))\}+w_i^{(m)}log\{S_u(Y_i|\*x_i,x_{si}(\cdot))\}.  \label{E2}
\end{eqnarray}
\subsection*{M-Step}
The M-step of the EM algorithm involves maximizing (\ref{E1}) and (\ref{E2}) with respect to the unknown parameters. Since (\ref{E1}) and (\ref{E2})  are separable functions of different parameters this can be performed separately for each set of parameters. We notice that maximizing (\ref{E1}) leads to the same likelihood as for a generalized scalar-on-function regression model with functional covariates \citep{reiss2017methods}. Moreover the cure submodel implies $g(\pi(\*z_i,z_{si}(\cdot)))=\*z_i^T\*b + \int_{0}^{1} z_{si}(s)b(s)ds$, where $g(\cdot)$ is a known link function. Hence we can use penalized spline based estimation approach \citep{marx1999generalized,wood2017generalized} for generalized additive model (GAM) for estimating the smooth coefficient function $b(s)$ and the parameter $\*b$. In particular, we use the \texttt{gam} function withing \texttt{mgcv} package \citep{wood2015package} in R which allows different link functions and the smoothing parameter can be calculated by the REML criterion.

For maximizing (\ref{E2}), we follow a penalized partial log-likelihood approach \citep{gellar2015cox,cui2021additive}. Following \cite{peng2000nonparametric,sy2000estimation}, we estimate $\bm\beta$ and $\beta(s)$ without specifying the baseline hazard function. We note that the conditional expectation (\ref{E2})  can be reformulated as,
\begin{eqnarray}
    E(l_{c_2}(\bm\beta,\beta(\cdot),S_0(t);\*O,\*B))=
    \sum_{i=1}^{n}\Delta_ilog\{w_i^{(m)}\lambda_u(Y_i|\*x_i,x_{si}(\cdot))\}+w_i^{(m)}log\{S_u(Y_i|\*x_i,x_{si}(\cdot))\}\notag\\
    =\sum_{i=1}^{n}\Delta_ilog\{w_i^{(m)}\lambda_0(Y_i) exp(\*x_i^T\bm\beta + \int_{0}^{1}x_{si}(s)\beta(s)ds)\}+w_i^{(m)}log\{S_0(Y_i)^{exp(\*x_i^T\bm\beta + \int_{0}^{1}x_{si}(s)\beta(s)ds)}\}\notag\\
    =log\prod_{i=1}^{n}[\lambda_0(Y_i) exp(log(w_i^{(m)})+\*x_i^T\bm\beta + \int_{0}^{1}x_{si}(s)\beta(s)ds)]^{\Delta_i}S_0(Y_i)^{exp(log(w_i^{(m)})+\*x_i^T\bm\beta + \int_{0}^{1}x_{si}(s)\beta(s)ds)}. \label{M2}
\end{eqnarray}

It can be noted that maximizing (\ref{M2}) is equivalent to maximizing a linear functional cox model \citep{gellar2015cox}. This can be achieved through maximizing the penalized partial log-likelihood. In particular, let the coefficient function $\beta(s)$ be modeled in terms of known basis function expansion such as $\beta(s)=\sum_{k=1}^{K}\theta_k B_k (s)$. Plugging in this expression in the latency submodel \ref{lat1}, we have  
\begin{eqnarray}
     log \lambda_{u}(t|\*x_i,x_{si}(\cdot))=log \lambda_{0}(t)+\*x_i^T\bm\beta + \sum_{k=1}^{K}\theta_k\int_{0}^{1}x_{si}(s)B_k(s)ds\notag\\
     =log \lambda_{0}(t)+\*x_i^T\bm\beta + \*v_i^T\bm\theta \notag
     =log \lambda_{0}(t)+\*U_i^T\bm\gamma. 
\end{eqnarray}
Here $\*U_i^T=(\*x_i^T,\*v_i^T)$, where $\*v_i^T$ is the vector with the elements $x_{si}(s)B_k(s)ds$  for $k=1,2,\ldots,K$ and $\bm\gamma=(\bm\beta^T,\bm\theta^T)^T$. The penalized partial loglikelihood is then given by 
\begin{eqnarray}
    l_{p}(\bm\gamma|\lambda)= l(\bm\gamma|\lambda)-\lambda P(\bm\theta)
=\sum_{i=1}^{n}\Delta_i[\*U_i^T\bm\gamma-log\sum_{Y_j\geq Y_i} e^{\{log(w_i^{(m)})+\*U_i^T\bm\gamma\}}]-\lambda P(\bm\theta). \label{ppl}
    \end{eqnarray}
In this article, we use cubic B-spline basis functions for modelling $\beta(s)$. However, other basis functions can also be used.
$P(\bm\theta)$ in (\ref{ppl}) is a roughness penalty on $\beta(s)$ and $\lambda$ is the corresponding smoothing parameter. We use a second order quadratic penalty and $P(\bm\theta)=\frac{1}{2}\bm\theta^T \^D \bm\theta$, where $\^D$ is a symmetric penalty matrix. For a fixed value of $\lambda$, we can use the Newton-Raphson algorithm for estimating the value of $\bm\gamma$ that maximizes the above penalized partial loglikelihood \citep{wood2016smoothing}. In particular, we use the \texttt{gam} function within \texttt{mgcv} package \citep{wood2015package} in R with the \texttt{cox.ph()} family for estimating the parameter $\bm\gamma$. The smoothing parameter $\lambda$ is obtained via maximization of the Laplace approximation
of the marginal likelihood of the smoothing parameter \citep{wood2016smoothing}.

\subsection*{Estimation of the Baseline Survival Function}
In order to proceed the E-step in the EM algorithm and using penalized partial likelihood for estimating the latency parameters, the estimated baseline survival function is updated using a Breslow-type estimator \citep{sy2000estimation}. The nonparametric estimator of the baseline survival function is given by,
$
    \hat{S}_0(t)=exp\{-\sum_{j:Y_{(j)}^*\leq t} \frac{D_{(j)}}{\sum_{i\in R_j} e^{\{log(w_i^{(m)})+\*U_i^T\bm\hat{\gamma}\}}}\},
$
where $Y_{(1)}^*<Y_{(2)}^*<\ldots<Y_{(k)}^*$ are the distinct and uncensored failure times, $D_{(j)}$ denote the number of events and $R_j$ is the risk set at time $Y_{(j)}^*$. Since $\hat{S}_0(t)$ may not go to $0$ as $t\rightarrow \infty$, we enforce the tail constraint $\hat{S}_0(t)=0$ for $t> Y_{(k)}^*$.
\subsection{Variance Estimation by Bootstrap}
The standard errors of the parameter estimates can be obtained using asymptotic variance based on the inverse of the observed information matrix corresponding to the
full data loglikelihood \citep{sy2000estimation} but this can be computationally complex especially as the number of functional covariates increase. In the context of functional data, we use a resampling subject-based bootstrap procedure \citep{crainiceanu2012bootstrap} for obtaining standard error and pointwise confidence intervals of the estimated parameters and coefficient functions in the proposed FPHMC model.

\section{Simulation study}
In this Section, we investigate the performance of the proposed FPHMC method using numerical simulations. To this end, we consider the following data-generating scenarios.
\subsection{Simulation Design}
We assume the functional logistic regression model 
\begin{equation*}
    log(\frac{\pi(\*z_i,z_{si}(\cdot))}{1-\pi(\*z_i,z_{si}(\cdot))})=b_0+z_{i1}b_1 +z_{i2}b_2 +\int_{0}^{1} z_{si}(s)b(s)ds,
\end{equation*}
for generating the cure probabilities. In the latency submodel,  the survival
times are generated from PH structure where the baseline survival time follows an exponential distribution with rate $exp(\beta_0)$, hence $log \lambda_{0}(t)=\beta_0$. Specifically, the latency submodel considered is 
\begin{equation*}
    log \lambda_{u}(t|\*x_i,x_{si}(\cdot))=log \lambda_{0}(t)+x_{i1}\beta_1+x_{i2}\beta_2 + \int_{0}^{1}x_{si}(s)\beta(s)ds.
\end{equation*}
Cured observations are associated with infinite survival times, and their survival times are set 
to a very large value (e.g., $10000$). Censoring times are generated from an exponential distribution with rate $\frac{1}{\mu_c}$, where $\mu_c$ (mean censoring time) is set to $10$ in all the scenarios considered. The scalar covariates in the latency submodel $x_{i1},x_{i2}$ are generated independently from a $Uniform(-1,1)$ distribution and for the cure submodel, $\*z$ is taken to be $\*z=(1,x_{1},x_{2})$. The functional covariate $x_{s}(\cdot)$ is generated as
$x_{si}(s)=\sum_{k=1}^{10}\psi_{ik}\phi_k(s)$, where $\phi_k(s)$ are orthogonal basis polynomials (of degree $k-1$) and $\psi_{ik}$ are  mean zero and independent Normally distributed scores with variance $\sigma^2_k=4(10-k+1)$. The functional covariate in the cure submodel $z_{s}(\cdot)$ is taken to be the same and equal to $x_{s}(\cdot)$. Following the above structure, three different data-generating scenarios are considered.

\subsection*{Scenario A: Moderate cure proportion}
In the cure submodel, the coefficient vector $\*b$ is taken to be $\*b=(1,2,0.5)$, and coefficient function is 
taken to be $b(s)=5sin (\pi s)$. This ensures a cure proportion around $37\%$. In the latency submodel the coefficient vector and coefficient functions are given by
$\bm\beta=(0.5,1)$, $\beta(s)=5cos (\pi s)$. The baseline log-hazard is taken as
$log\lambda_0(t)=\beta_0=0.5$.
\subsection*{Scenario B: Low cure proportion}
In this scenario, the coefficient vector $\*b$  in the cure submodel is taken to be $\*b=(3,2,0.5)$, and the coefficient function is 
taken to be $b(s)=5sin (\pi s)$. This ensures a cure proportion around $16\%$. In the latency submodel the coefficient vector $\bm\beta$, coefficient function $\beta(s)$ and the baseline log-hazard $\beta_0$ are kept the same as in scenario A.
\subsection*{Scenario C: High cure proportion}
In this scenario, the coefficient vector $\*b$  in the cure submodel is taken to be $\*b=(-1.5,-2,-0.5)$, and coefficient function is 
taken to be $b(s)=5sin (\pi s)$. This achieves a cure rate of around $69\%$. In the latency submodel the coefficient vector, coefficient functions and the baseline log-hazard are again kept the same as in scenario A.

For all the above scenarios, three sample sizes $n\in\{300,500,1000\}$ are considered. We generate 100 Monte-Carlo (M.C) replications from the above data-generating scenarios to assess the performance of the proposed FPHMC method.

\subsection{Simulation Results}
\subsection*{Performance under scenario A}
\begin{figure}[H]
	\centering
 \begin{subfigure}[b]{0.45\textwidth}
         \centering
        \includegraphics[width=0.99\textwidth,height=0.82\textwidth]{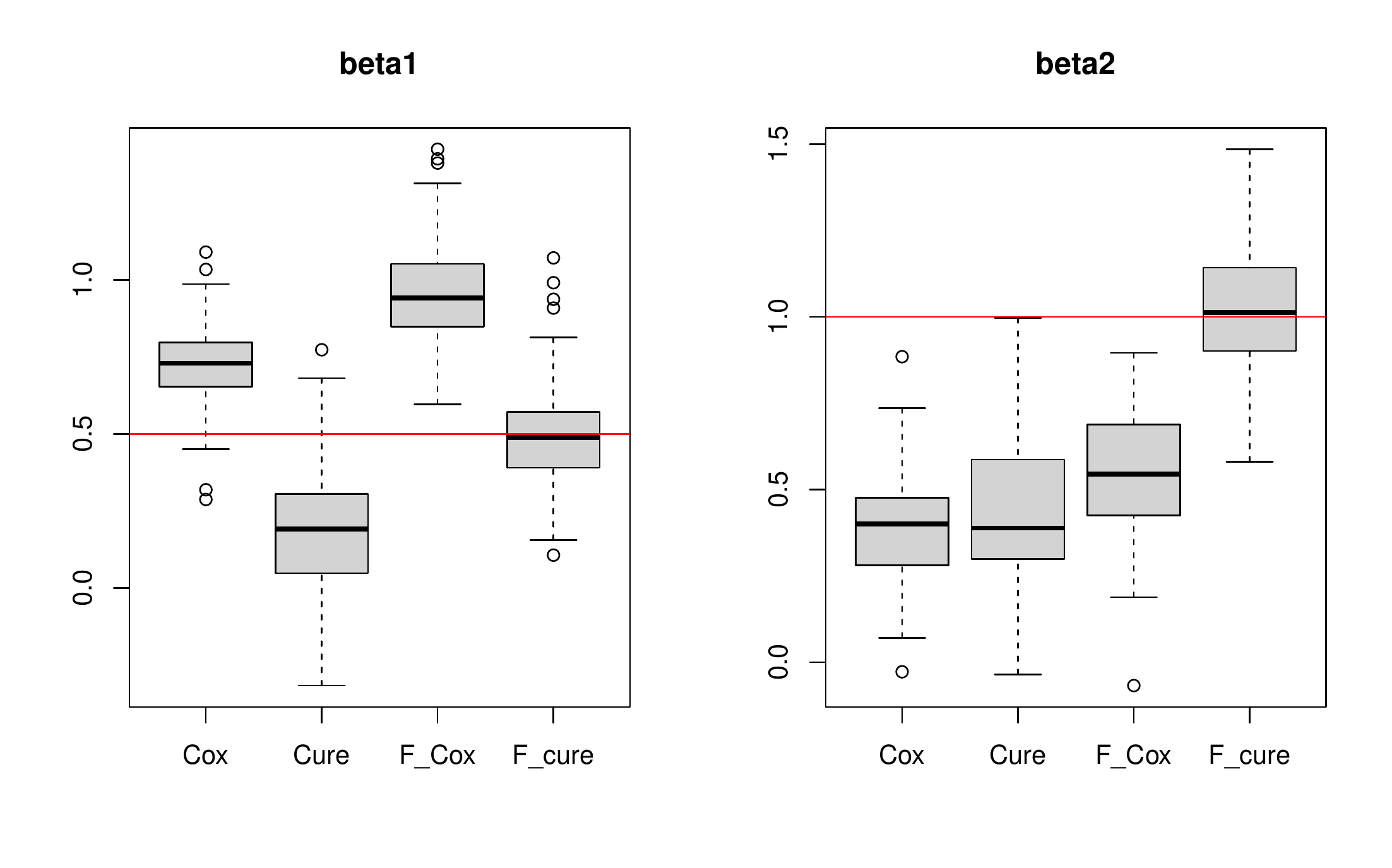}
\caption{Boxplots of $\hat{\beta}_1,\hat{\beta}_2$ from the Cox model and mixture cure model (Cure) model with scalar covariates, linear functional cox model (F-Cox) and the proposed FPHMC method (F-cure), scenario A, n=300. The solid red line indicates the true value of the parameters.} 
\label{fig:t1}	
     \end{subfigure}
     \hfill
     \begin{subfigure}[b]{0.45\textwidth}
        \centering\includegraphics[width=0.99\textwidth]{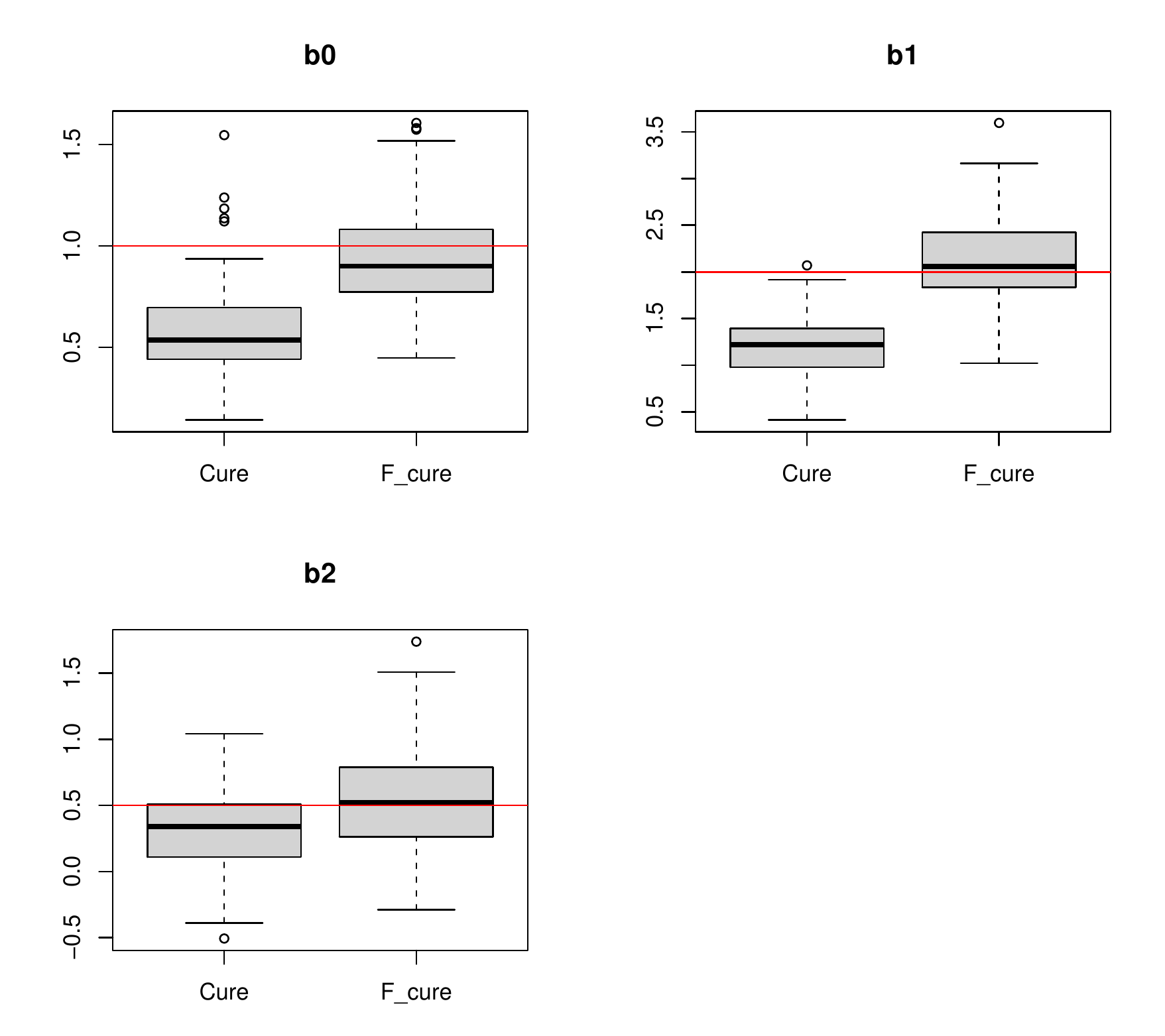}
\caption{Boxplots of $\hat{b}_0,\hat{b}_1,\hat{b}_2$ from the mixture cure model (Cure) model with scalar covariates and the proposed FPHMC method (F-cure), scenario A, n=300. The solid red line indicates the true value of the parameters.}
         \label{fig:t2}
     \end{subfigure}
    \vfill
     \begin{subfigure}[b]{0.45\textwidth}
         \centering
        \includegraphics[width=0.99\textwidth,height=0.82\textwidth]{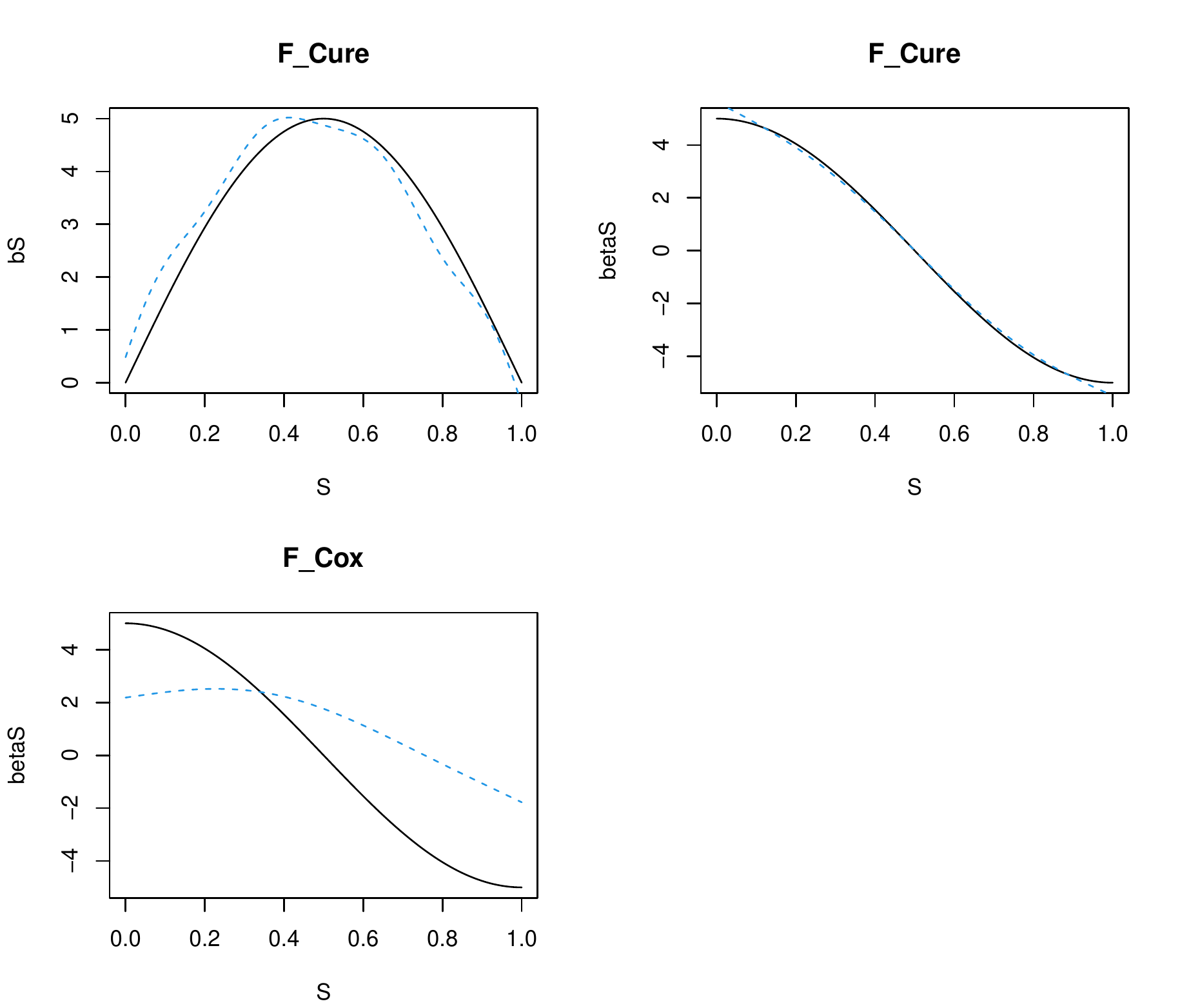}
\caption{Top left: True incidence coefficient function $b(s)$ (solid) and average estimated $\hat{b}(s)$ (dotted) from the FPHMC method, Top right: True latency coefficient function $\beta(s)$ (solid) and average estimated $\hat{\beta}(s)$ (dotted) from the FPHMC method, Bottom left: True latency coefficient function $\beta(s)$ (solid) and average estimated $\hat{\beta}(s)$ (dotted) from the linear functional Cox method, scenario A, n=300. All estimated functions are averaged over 100 M.C replications } 
\label{fig:t3}	
     \end{subfigure}
     \hfill
      \begin{subfigure}[b]{0.45\textwidth}
         \centering
        \includegraphics[width=0.99\textwidth,height=1\textwidth]{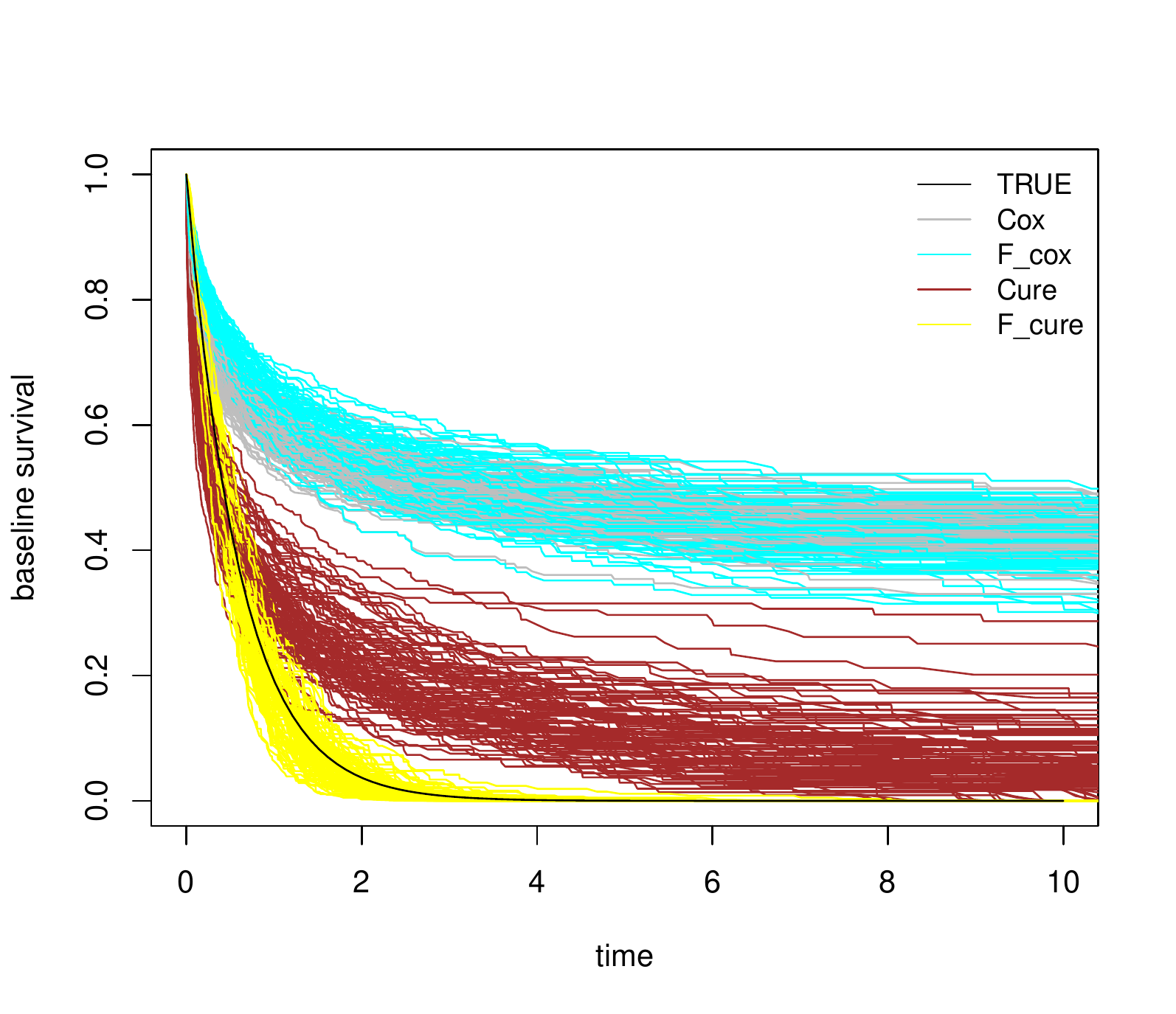}
\caption{Displayed are the true baseline survival function (black curve) and estimates of the baseline survival function $S_0(t)$ from the Cox model and mixture cure model (Cure) model with scalar covariates, linear functional cox model (F-Cox) and the proposed FPHMC method (F-cure), scenario A, n=300.} 
\label{fig:t4}	
     \end{subfigure}
\caption{Simulation Results, scenario A, n=300.}
\end{figure}

We apply the proposed FPHMC method to estimate the cure and latency model parameters $(b_0,b_1,b_2,\beta_1,\beta_2)$, the coefficient functions $b(s)$ and $\beta(s)$ and the baseline survival function $S_0(t)$. For comparison purposes and to illustrate the gain using the proposed method with functional covariates, we also fit the cox model with scalar covariates, the mixture cure model with scalar covariates, and the linear functional cox model \citep{gellar2015cox}. Figure \ref{fig:t1} displays the boxplot of the distribution of the estimated parameters $\hat{\beta}_1,\hat{\beta}_2$ from the FPHMC method and the three competing approaches for sample size $n=300$.
We observe that the estimates $\hat{\beta}_1,\hat{\beta}_2$ from the cox model, mixture cure model and linear functional cox model are highly biased, the interquartile interval failing to contain the true parameter value. On the other hand, the proposed FPHMC method is seen to produce accurate estimates of these parameters. The estimates of the cure model parameters $(b_0,b_1,b_2)$ are available from the mixture cure model and proposed FPHMC method, and their distribution is shown in Figure \ref{fig:t2}. We again observe the proposed FPHMC method yielding more accurate and less biased estimates compared to the usual mixture cure model. The estimates of the functional coefficients $b(s)$ and $\beta(s)$ from the FPHMC method averaged over 100 M.C replications are shown in Figure \ref{fig:t3}. The estimated coefficient function for $\beta(s)$ from the linear functional cox model is also included for comparison.
The FPHMC method can be seen to capture the coefficient function $\beta(s)$ more accurately. We also report
the integrated mean squared error (MSE), integrated squared Bias (Bias$^2$), and integrated variance (Var) of the estimates of all the functional parameters  ($b(s),\beta(s)$) for the FPHMC method in Table \ref{tab:my-table11} across all three sample sizes. For a functional effect $\beta(s)$, these quantities are defined as $MSE=\frac{1}{M} \sum_{j=1}^{M}\int_{0}^{1}\{\hat{\beta}^{j}(s)-\beta(s)\}^2ds$, Bias$^2=\int_{0}^{1}\{\hat{\bar{\beta}}(s)-\beta(s)\}^2dp$, $Var=\frac{1}{M} \sum_{j=1}^{M}\int_{0}^{1}\{\hat{\beta}^{j}(s)-\hat{\bar{\beta}}(s)\}^2ds$. 

Here $\hat{\beta}^{j}(s)$ is the estimate of $\beta(s)$ from the $j$th replicated dataset and $\hat{\bar{\beta}}(s)=\frac{1}{M} \sum_{j=1}^{M}\hat{\beta}^{j}(s)$ is the M.C average estimate based on the M replications. MSE as well as squared Bias and Variance  are found to decrease for both $b(s)$ and $\beta(s)$ and become negligible as sample size $n$ increases, illustrating the satisfactory accuracy of the proposed estimator.
\begin{table}[H]
\centering
\caption{Integrated squared bias, variance and mean square error  of estimated $b(s)$ and $\beta(s)$ over 100 Monte-Carlo replications, Scenario A. }
\label{tab:my-table11}
\begin{tabular}{lllllll}
\hline
Parameter              & \multicolumn{3}{l|}{$b(s)$}                                                       & \multicolumn{3}{l}{$\beta(s)$}                                                       \\ \hline
Sample Size & \multicolumn{1}{l}{Bias$^2$}              & \multicolumn{1}{l}{Var}    & MSE    & \multicolumn{1}{l}{Bias$^2$}              & \multicolumn{1}{l}{Var}    & MSE    \\ \hline
n= 300                   & \multicolumn{1}{l}{0.15}               & \multicolumn{1}{l}{2.53} & 2.68 & \multicolumn{1}{l}{$0.03$} & \multicolumn{1}{l}{0.19} & 0.22 \\ \hline
n= 500                   & \multicolumn{1}{l}{$0.07$}               & \multicolumn{1}{l}{1.11} & 1.18 & \multicolumn{1}{l}{$0.01$} & \multicolumn{1}{l}{0.11} & 0.12 \\ \hline
n= 1000                   & \multicolumn{1}{l}{$0.02$} & \multicolumn{1}{l}{0.45} & 0.47 & \multicolumn{1}{l}{$0.004$} & \multicolumn{1}{l}{0.05} & 0.05 \\ \hline
\end{tabular}
\end{table}

The estimated baseline survival function $\hat{S}_0(t)$ from the FPHMC method across all the replications along with the true baseline survival function $S_0(t)$ are displayed in 
Figure \ref{fig:t4}, for the case $n=300$. The estimated baseline survival functional from the other three competing methods are also shown. It can be clearly observed, apart from the proposed FPHMC method all the approaches clearly overestimate the true baseline survival function. The proposed FPHMC method estimates the baseline survival function accurately and also appears to have lower variability. The performance of the estimators for other sample sizes are similar and are omitted for brevity. 
The above results clearly illustrate the satisfactory performance of the FPHMC method in the presence of a cure fraction and functional covariates. 

The results under scenario B and Scenario C are reported in the Appendix A of the Supplementary Material which show robust performance of the proposed method in the presence of low and high cure proportions.

\section{Real data examples}
\subsection{NHANES 2003-2006}
We apply the proposed FPHMC method to data from the NHANES waves 2003–2006. NHANES data can be linked to the National Death Index (NDI) \citep{leroux2019organizing} for collecting mortality information. In particular, we use the 2019 (December 31) mortality information from NDI (\url{https://www.cdc.gov/nchs/data-linkage/mortality-public.htm}) to define our survival outcome. The NHANES aims to provide a broad range of descriptive health and nutrition statistics for the civilian non-institutionalized population of the U.S. \citep{johnson2014national}. Data collection consists of an interview and an examination; the interview gathers person-level demographic, health, and nutrition information; the examination includes physical measurements, such as blood pressure, a dental examination, and the collection of blood and urine specimens for laboratory testing. Additionally, NHANES 2003-06 provide objectively measured physical activity data collected by hip-worn accelerometer. The participants were asked to wear a physical activity monitor, starting on the day of their exam, and to keep wearing this device all day and night for seven full days (midnight to midnight) and remove it on the morning of the ninth day. The device used was the Actigraph AM-7164 (Actigraph, Ft. Walton Beach, FL) uni-axial accelerometer \citep{varma2017re}.

A total of 2816 adults aged 50–85 years (with physical activity monitoring available at least ten hours per day for at least four days) with available mortality and covariate information were included in this analysis \citep{cui2021additive}. Here for each participant $i=1,\dots,n$, we observe physical activity $X_{ij}(t)$, $t\in [0,1440]$, $j=1,\dots,n_i$ which denote the physical activity (log-transformed $A\rightarrow log(1+A)$) for the individual $i$ on the day $j$ at time $t$. To summarize the physical activity (PA) information for each participant in a common functional profile, we consider the diurnal average functional curve  $X_i(t)= \frac{1}{n_i} \sum_{j=1}^{n_i} X_{ij}(t)$ as the functional covariate of interest. The diurnal curves $X_i(t)$ are further aggregated into 10 minutes epochs to make the PA profiles smooth for each subject. 

Survival time is measured in months from accelerometer
wear end and all subjects are censored on December 31, 2019 based on the mortality information reported in NDI 2019 release. Among the 2816 study participants considered at the baseline, 256 ($9.1\%$) were deceased due to cancer. The objective of this analysis is to study the association between the baseline diurnal patterns of physical activity and all-cancer mortality while adjusting for other biological factors. For this cause-specific analysis, the study participants with death due to reasons other than Cancer ($N=861$) were considered right-censored. As observed in Figure \ref{fig:kmage1}, there seems to be a cure fraction in the population, subjects who will not experience death due to cancer.
Hence using the proposed FPHMC model would be more suitable for this situation.

We apply the proposed FPHMC method to model the effect of daily PA on all cancer mortality while adjusting for the scalar latency and cure  covariates listed in Table \ref{tab:regscalar}. Briefly, the scalar covariates include demographic information (age, gender, BMI, race) and biological factors (smoking and cancer). The diurnal PA profile is included as a functional covariate in the latency submodel. From a practical point of view, the demographic and biological covariates control the probability of risk (cure), whereas the biological covariates along with daily physical activity levels could be important risk factors related to all-cancer mortality. We apply $B=1000$ bootstrap to quantify uncertainty of the estimated parameters. Table \ref{tab:regscalar} reports the estimated scalar coefficients of the cure and latency submodel in the forms of hazard ratios for the latency submodel and odds ratio (odds ratio for being susceptible) for the cure submodel along with their $95\%$ confidence intervals. Age is found to be a critical factor in assigning subjects to the susceptible group for cancer (Odds ratio equal to 1.07). Smoking is also found to be a significant risk factor (odds ratio 2.5 for former and 3.6 for current smokers) for being susceptible to cancer. In the latency submodel, females are found to have lower risk of cancer morality compared to males (hazard ratio 0.41) after adjusting for the other scalar covariates and diurnal PA profile. Having cancer at the baseline is found to be associated with increased risk of cancer mortality (hazard ratio 3.02). 
\begin{table}[H]
\centering
\caption{Estimated Scalar coefficients (as odds ratio of being susceptible and hazard ratio ) of the cure and latency submodels from the FPHMC method along with their $95\%$ confidence intervals in the NHANES application. The scalar covariates included are $Age$, $Gender:$  Binary variable that indicates if the subject is male (reference) or female. $Race:$   A categorical variable indicates if the subject belongs to the following groups: i)	White (reference) ($58.9\%$); ii) Mexican American ($18.1\%$); iii) Other Hispanic ($1.9\%$), iv) Black ($18.3\%$) v) Other Race ($2.7\%$). 
$Smoke:$  Categorical variable with the following levels i) Never (reference), ii) Former and iii) Current.
$Cancer:$  Binary variable indicating if the subject had cancer.
} 
\label{tab:regscalar}
\begin{tabular}{l|c|c}
  \hline
 Variables  & Odds ratio (Cure part) & Hazard ratio (Latency part)  \\ 
  \hline
Age  & 1.07 (1.04,1.10) & 0.99 (0.96,1.02)\\ 
  Gender Female & 0.96 (0.54,1.69) & 0.41 (0.19,0.88)  \\ 
  BMI & 1.00 (0.96,1.04) & 1.00 (0.96,1.05) \\
  Race Mexican American & 0.74 (0.36,1.51) & 1.65 (0.65,4.16)   \\ 
  Race Other Hispanic   & 0.29 (1.2$\times10^{-4}$,6.6$\times10^{2}$) & 2.94 (1.7$\times10^{-18}$,5.1$\times10^{18}$) \\ 
  Race Black & 1.50 (0.89,2.53) & 1.22 (0.66,2.27)   \\ 
  Race Other & 0.97 (0.06,14.80)  & 0.68 (3.6$\times10^{-6}$,1.3$\times10^{5}$)   \\ 
  Smoke Former   & 2.53 (1.41,4.55) & 0.71 (0.35,1.45)  \\ 
  Smoke Current & 3.59 (1.86,6.92) & 1.03 (0.43,2.45) \\ 
  Cancer & 1.14 (0.64,2.04) & 3.02 (1.59,5.75)  \\ 
  \hline
\end{tabular}
\vspace{6 mm}
\end{table}

\begin{figure}[H]
\centering\includegraphics[width=0.7\textwidth]{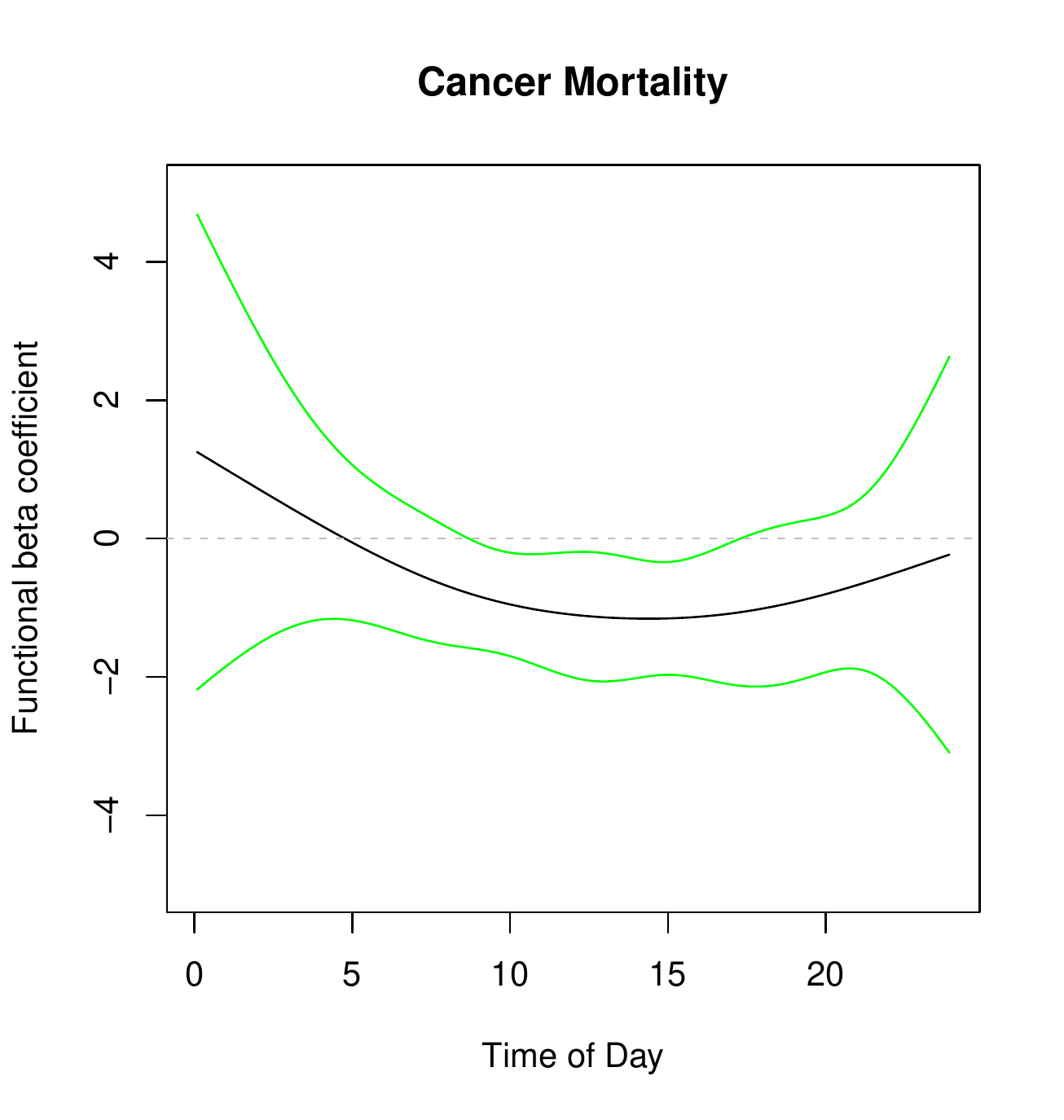}
\caption{Estimated functional effect $\beta(s)$ of baseline diurnal PA in NHANES 2003-06 on log hazard of cancer mortality (2019) from the FPHMC model along with its $95\%$ point-wise confidence interval.}
\label{fig:coefficient}
\end{figure}

Figure \ref{fig:coefficient} displays the estimated functional-coefficient $\beta(s)$ from the latency component of the FPHMC model. It captures the impact of daily physical activity patterns on the log-hazard of all cancer mortality among the susceptibles after adjusting for age and other confounders. We observe that an increase in physical activity during the day (8 a.m - 6 p.m.) is associated with a reduced hazard of cancer mortality after adjusting for all the other variables.
This period primarily corresponds to the time-of-day where humans expend more energy. Interestingly, this protective effect is maintained until night/midnight but is not statistically significant. In the late-night period, the accelerometer monitor generally takes values zero; consequently, the contribution of PA to the overall risk is minimal. 

Our results using the FPHMC model complement the findings from other physical activity research about the benefits of being active during the day in longevity. Previous research in this direction focused on predicting all-cause or CVD (cardiovascular) mortality to five or ten years in NHANES 2003-2006 \citep{10.1093/gerona/glz193,cui2021additive,ledbetter2022cardiovascular} or in describing variations for different age-group or other medical conditions \citep{wrobel2021diurnal, keller2022using, mcdonnell2021registration}.
Our findings is one of the first to show association between daily patterns of PA and cancer mortality \citep{schmid2014association,patel2015leisure,liu2016leisure} after adjusting for age and other confunders, which could be very important for designing time-of-day specific PA interventions.

\subsection{Post-ICU Mortality in ICAP}
We use intensive care unit (ICU) data from  the Improving Care of Acute Lung Injury Patients (ICAP) study \citep{needham2005study} to examine the association between daily functional measures of organ function status collected in the intensive care unit and patient recovery and survival. \cite{gellar2014variable,gellar2015cox} have previously analyzed this data  using a linear functional cox model to study the effect of patients' overall organ function status in the ICU on post-ICU mortality. The ICAP study is a prospective cohort study examining the long-term outcomes of patients suffering from acute lung injury or acute respiratory distress syndrome (ALI/ARDS). Further details about the ICAP study can be found in \cite{needham2005study,gellar2015cox}. 

The original ICAP study enrolled 520 subjects among which there were 237 (46\%) deceased in the ICU. Among the 283 survivors, 16 subjects (5.7\%) did not consent to follow-up and were excluded from this analysis.
The remaining 267 subjects were followed for up to two years from their date of enrollment and mortality information was recorded to the nearest day based on the family reports or publicly available records. In the ICAP study, we have daily measurements of patients' Sequential Organ Failure Assessment (SOFA) score during their ICU stay. SOFA score serves as a composite score of a patient's overall organ function status in the ICU consisting of six physiological components with higher scores representative of poorer organ function. The total SOFA score is aggregated across the components and ranges from $0–24$. We consider each subject's history of measured SOFA scores over time a functional covariate, $X_i(t)$, where $t\in [0, T_i]$ is the ICU day. As the random process can have different duration, we consider the following domain-transformed SOFA score \citep{gellar2015cox} $\widetilde{X}_i(s)= X_{i}(sT_i)$, where $s:=t/T_i$, that are each defined over $[0, 1]$. 

Our goal is to investigate the association between post-hospital mortality and a patient's SOFA function while adjusting for the other scalar covariates: age, gender, and Charlson co-morbidity index (a measure of baseline health) \citep{charlson1987new}. We consider “time zero” to be the day the subject is discharged from the hospital following ALI/ARDS, and subjects are censored at two years following their diagnosis. Among the 267 patients considered, 90 ($33.7\%$) were deceased at the time of censoring. Supplementary Figure S9 displays the marginal Kaplan Meier survival estimate which seems to follow a cured structure. A fraction of discharged patients might be at a relatively lower risk in the time period considered (2 years) while another fraction of the patients might be susceptible, whose survival might depend upon age, existing health conditions and baseline organ function status represented by the SOFA score. 
 
We apply the proposed FPHMC method in this paper with age, gender, Charlson co-morbidity index and SOFA score as the covariates in both the cure and latency submodels. We perform B = 1000 bootstrap to quantify uncertainty of the estimated parameters. Table \ref{tab:coefuci} shows the estimated coefficients for the scalar variables in forms of odds ratio (for cure submodel) and hazard ratio (for latency submodel). In the cure part, age (odds ratio 1.06) and the Charlson co-morbidity index (odds ratio 1.34) are found to be positively associated with a higher risk of being susceptible. For the latency part, none of the variables achieves statistical significance. The SOFA score is considered as a functional covariate in both the cure and latency submodels and is not found to be significant in either of them, which matches with the findings of \cite{gellar2015cox}, where the SOFA score was not found be associated with post-ICU mortality. Supplementary Figure S10 displays the estimated functional coefficients $b(s)$ and $\beta(s)$ of the SOFA score in the cure and latency submodel respectively.
In the future, more powerful functional biomarkers might be required to characterize long-term patient survival. 

\begin{table}[H]
\centering
\caption{Estimated Scalar coefficients (odds ratio and hazard ratio) of the cure and latency submodels from the FPHMC method along with their $95\%$ confidence intervals in the ICAP application.}
\label{tab:coefuci}
\begin{tabular}{l|rrrrrrr}
  \hline
 Variables & Odds ratio (Cure part) & Hazard ratio (Latency part)  \\ 
  \hline
age & 1.06 (1.04,1.09)& 1.00 (0.98,1.03)  \\ 
  Gender (Male)  & 1.43 (0.78.2.62)  & 0.62 (0.34,1.13) \\ 
  Charlson Index  & 1.34 (1.18,1.53) & 1.05 (0.97,1.13)\\ 
   \hline
\end{tabular}
\vspace{0.5cm}
\end{table}

\section{Discussion}
The paper's primary contribution is to propose the first functional proportional hazard mixture cure (FPHMC) model for right-censored data that involves the simultaneous incorporation of smoothness in the estimation of the functional coefficients. The new method is based on solving the optimization problem with the EM algorithm and semiparametric penalized spline-based estimation of the model coefficients. The proposed method is also computationally efficient, allowing resampling techniques to obtain point-wise confidence intervals simply using the Gaussian asymptotic approximation of the model coefficients. The theoretical properties of the underlying cure and latency submodels: i) the generalized scalar on function regression model; ii) the linear functional Cox model guarantees the good model behavior of this mixture regression model. 
In the cure survival analysis literature, the only functional model has been explicitly proposed for interval-censored data \cite{shi2022}. Nevertheless, it does not involve any smoothing process of the coefficients, which, as we know from previous functional data analysis literature, can be a critical problem in functional data analysis, especially in contaminated and noisy data settings.

Numerical analysis using simulations
have illustrated the satisfactory finite-sample performance of the proposed FPHMC method in accurately estimating the scalar, functional coefficients and the baseline survival function, in the presence of a cure fraction. Traditionally, cure models have been used in medical oncology problems \citep{felizzi2021mixture,chen2013statistical}, as in the case of immunotherapy studies \citep{wei2020cancer}. Here, we have explored new applications of cure models, such as in the analysis of patients in the intensive care unit or the impact of physical activity on cancer mortality. Importantly, here we introduce high-resolution information on the physiological processes of patients by incorporating biological signals as functional covariates. In the NHANES 2003-06 application, our results demonstrate the association between increased physical activity during the central hours of the day and reduced risk of cancer mortality using objectively measured PA data, complementing findings of previous studies which primarily used various summary measures of PA \citep{liu2016leisure,patel2019american}. While in the case of ICAP study, we did not find any statistically significant effect of SOFA score on post-ICU mortality, indicating that this score may not reflect the future evolution of the patients. In order to assign the patient to a risk or non-risk group, age and Charlson index are found to be the relevant factors.


This current work has opened up possibilities of multiple research directions. In future work, we will propose specific goodness-of-fit procedures for cure models, methods to handle distributional physical activity representations \citep{doi:10.1177/0962280221998064,matabuena2022physical,Ghosal2022} which are compositional functional objects.
New cure regression models based on the RKHS statistical learning paradigm with the potential to handle complex statistical predictors as graphs, functional data and probability distributions in a natural way could be explored. Another possible direction would be to 
extend the proposed method to multilevel functional data \citep{goldsmith2015generalized} which could be directly applicable for example to PA data from NHANES. 
In this article, we have used a subject-level bootstrap for uncertainty quantification, other resampling strategies such as the wild bootstrap could be explored. In addition, an important research direction is the pursuit of prediction intervals using uncertainty quantification methods that guarantee finite sample coverages, such as conformal inference techniques \citep{candes2021conformalized}. New methods have emerged in the case of right-censoring \citep{candes2021conformalized,teng2021t,gui2022conformalized}, but to the best of our knowledge, there is yet to be a proposal for cure models and functional data.    


\section*{Supplementary Material}
Appendix A, Supplementary Tables S1-S2 and Supplementary Figures S1-S10 are available with this paper as Supplementary Material.
\section*{Software}
R software implementation and illustration of the proposed method is available with this paper and will be made publicly available on Github.

\bibliographystyle{Chicago}
\bibliography{refs}
\end{document}


\def\spacingset#1{\renewcommand{\baselinestretch}%
{#1}\small\normalsize} \spacingset{1}


\if0\blind
{
  \title{\bf Supplementary Material for Functional proportional hazards mixture cure model and its application to modelling the association between cancer mortality and physical activity in NHANES 2003-2006 }
 \author{Rahul Ghosal$^{1,\ast}$, Marcos Matabuena$^{2}$, Jiajia Zhang$^{3}$ \\
  \\
$^{1}$ Department of Epidemiology and Biostatistics, University of South Carolina \\
$^{2}$ Centro Singular de Investigación en Tecnologías Intelixentes, \\Universidad de Santiago de Compostela, Santiago de Compostela, Spain\\
$^{3}$ Department of Epidemiology and Biostatistics, University of South Carolina\\
}
  \maketitle
} \fi

\if1\blind
{
  \bigskip
  \bigskip
  \bigskip
  \begin{center}
    {\LARGE\bf Title}
\end{center}
  \medskip
} \fi

\bigskip

\vfill

\newpage
\spacingset{1.5} 

\section{Appendix A: Simulation Scenarios}
\subsection*{Performance under scenario B}
We apply the proposed FPHMC method along with the three competing methods as in scenario A to estimate the model parameters $(b_0,b_1,b_2,\beta_1,\beta_2)$, coefficient functions $b(s),\beta(s)$ and the baseline survival function $S_0(t)$. The boxplot of the distribution of $\hat{\beta}_1,\hat{\beta}_2$ are shown in Supplementary Figure S1 for sample size $n=300$. Supplementary Figure S2 displays the distribution of the estimated cure model parameters $(\hat{b}_0,\hat{b}_1,\hat{b}_2)$. Overall, we observe that the proposed FPHMC method yield more accurate and less biased estimates of both cure and latency model parameters compared to the competing methods. Supplementary Figure S3 displays the estimated functional coefficients $\hat{b}(s),\hat{\beta}(s)$ from the FPHMC method averaged over 100 M.C replications. The estimated $\hat{\beta}(s)$ seem to be capturing the true latency coefficient function $\beta(s)$ accurately. The integrated mean squared error (MSE), integrated squared Bias (Bias$^2$) and integrated variance (Var) for the functional parameters ($b(s),\beta(s)$) are reported in Supplementary Table S1 across all three sample sizes. The estimated $\hat{b}(s)$ can be noticed to be more biased compared to scenario A, and this might be attributed to the low cure proportion in this scenario. Supplementary Figure S4 displays the estimated baseline survival function $\hat{S}_0(t)$ using the FPHMC method and the other three competing methods for $n=300$. We can again observe a superior performance of the FPHMC method in terms of lower bias and variability. The performance of the estimator for the other sample sizes are similar and are excluded due to conciseness.

\subsection*{Performance under scenario C}
The cure proportion in this scenario is relatively high ($\sim 69\%$). The FPHMC model and the three competing methods are applied to estimate the scalar model parameters, coefficient functions and the baseline survival function.  Supplementary Figure S5 and Supplementary Figure S6 shows the distribution of $\hat{\beta}_1,\hat{\beta}_2$ and cure model parameters $(\hat{b}_0,\hat{b}_1,\hat{b}_2)$ for sample size $n=300$. The proposed FPHMC method can be seen to capture the parameters accurately and with negligible bias compared to the other approaches. Supplementary Figure S7 displays the estimated functional coefficients $\hat{b}(s),\hat{\beta}(s)$ from the FPHMC method averaged over 100 M.C replications. The integrated mean squared error (MSE), integrated squared Bias (Bias$^2$) and integrated variance (Var) for the functional parameters ($b(s),\beta(s)$) are reported in Supplementary Table S2 across all three sample sizes. We observe that even in the presence of high cure proportion, the proposed FPHMC method estimates the coefficient functions accurately.
The estimated baseline survival function $\hat{S}_0(t)$ using the FPHMC method and the other three competing approaches are shown in Supplementary Figure S8 for $n=300$. The FPHMC method is observed to capture the true baseline survival function with lower bias and variability.
The scalar cox model and linear functional cox model do not adjust for cure proportion and can be noticed to be highly biased. The performance of the estimator for the other sample sizes are similar and are excluded for brevity.

\section{Supplementary Tables}

\begin{table}[H]
\centering
\caption{Integrated squared bias, variance and mean square error  of estimated $b(s)$ and $\beta(s)$ over 100 Monte-Carlo replications, Scenario B. }
\label{tab:my-table}
\begin{tabular}{lllllll}
\hline
Parameter              & \multicolumn{3}{l|}{$b(s)$}                                                       & \multicolumn{3}{l}{$\beta(s)$}                                                       \\ \hline
Sample Size & \multicolumn{1}{l}{Bias$^2$}              & \multicolumn{1}{l}{Var}    & MSE    & \multicolumn{1}{l}{Bias$^2$}              & \multicolumn{1}{l}{Var}    & MSE    \\ \hline
n= 300                   & \multicolumn{1}{l}{0.33}               & \multicolumn{1}{l}{4.33} & 4.66 & \multicolumn{1}{l}{$0.02$} & \multicolumn{1}{l}{0.13} & 0.15 \\ \hline
n= 500                   & \multicolumn{1}{l}{$0.205$}               & \multicolumn{1}{l}{2.11} & 2.31 & \multicolumn{1}{l}{$0.008$} & \multicolumn{1}{l}{0.07} & 0.08 \\ \hline
n= 1000                   & \multicolumn{1}{l}{$0.07$} & \multicolumn{1}{l}{0.93} & 1.00 & \multicolumn{1}{l}{$0.003$} & \multicolumn{1}{l}{0.04} & 0.04 \\ \hline
\end{tabular}
\end{table}

\begin{table}[H]
\centering
\caption{Integrated squared bias, variance and mean square error  of estimated $b(s)$ and $\beta(s)$ over 100 Monte-Carlo replications, Scenario C. }
\label{tab:my-table}
\begin{tabular}{lllllll}
\hline
Parameter              & \multicolumn{3}{l|}{$b(s)$}                                                       & \multicolumn{3}{l}{$\beta(s)$}                                                       \\ \hline
Sample Size & \multicolumn{1}{l}{Bias$^2$}              & \multicolumn{1}{l}{Var}    & MSE    & \multicolumn{1}{l}{Bias$^2$}              & \multicolumn{1}{l}{Var}    & MSE    \\ \hline
n= 300                   & \multicolumn{1}{l}{0.09}               & \multicolumn{1}{l}{1.98} & 2.07 & \multicolumn{1}{l}{$0.07$} & \multicolumn{1}{l}{0.37} & 0.44 \\ \hline
n= 500                   & \multicolumn{1}{l}{$0.05$}               & \multicolumn{1}{l}{1.25} & 1.30 & \multicolumn{1}{l}{$0.05$} & \multicolumn{1}{l}{0.22} & 0.27 \\ \hline
n= 1000                   & \multicolumn{1}{l}{$0.01$} & \multicolumn{1}{l}{0.59} & 0.60 & \multicolumn{1}{l}{$0.01$} & \multicolumn{1}{l}{0.13} & 0.14 \\ \hline
\end{tabular}
\end{table}

\section{Supplementary Figures}

\begin{figure}[H]
\centering
\includegraphics[width=1\linewidth , height=.6\linewidth]{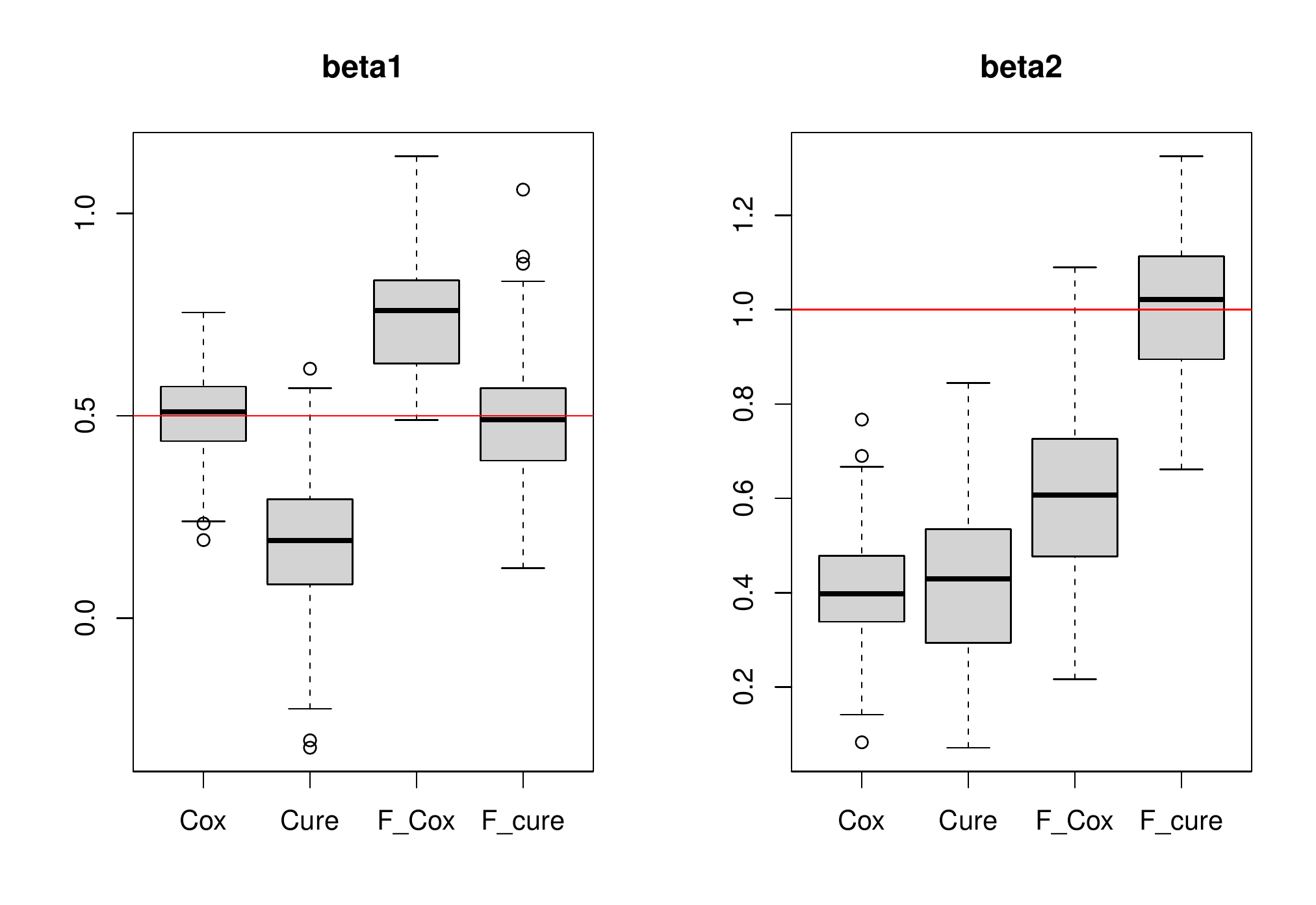}
\caption{Boxplots of $\hat{\beta}_1,\hat{\beta}_2$ from the Cox model and mixture cure model (Cure) model with scalar covariates, linear functional cox model (F-Cox) and the proposed FPHMC method (F-cure), scenario B, n=300. The red solid line indicates the true value of the parameters.}
\label{fig:fig2b}
\end{figure}

\begin{figure}[H]
\centering
\includegraphics[width=1\linewidth , height=1\linewidth]{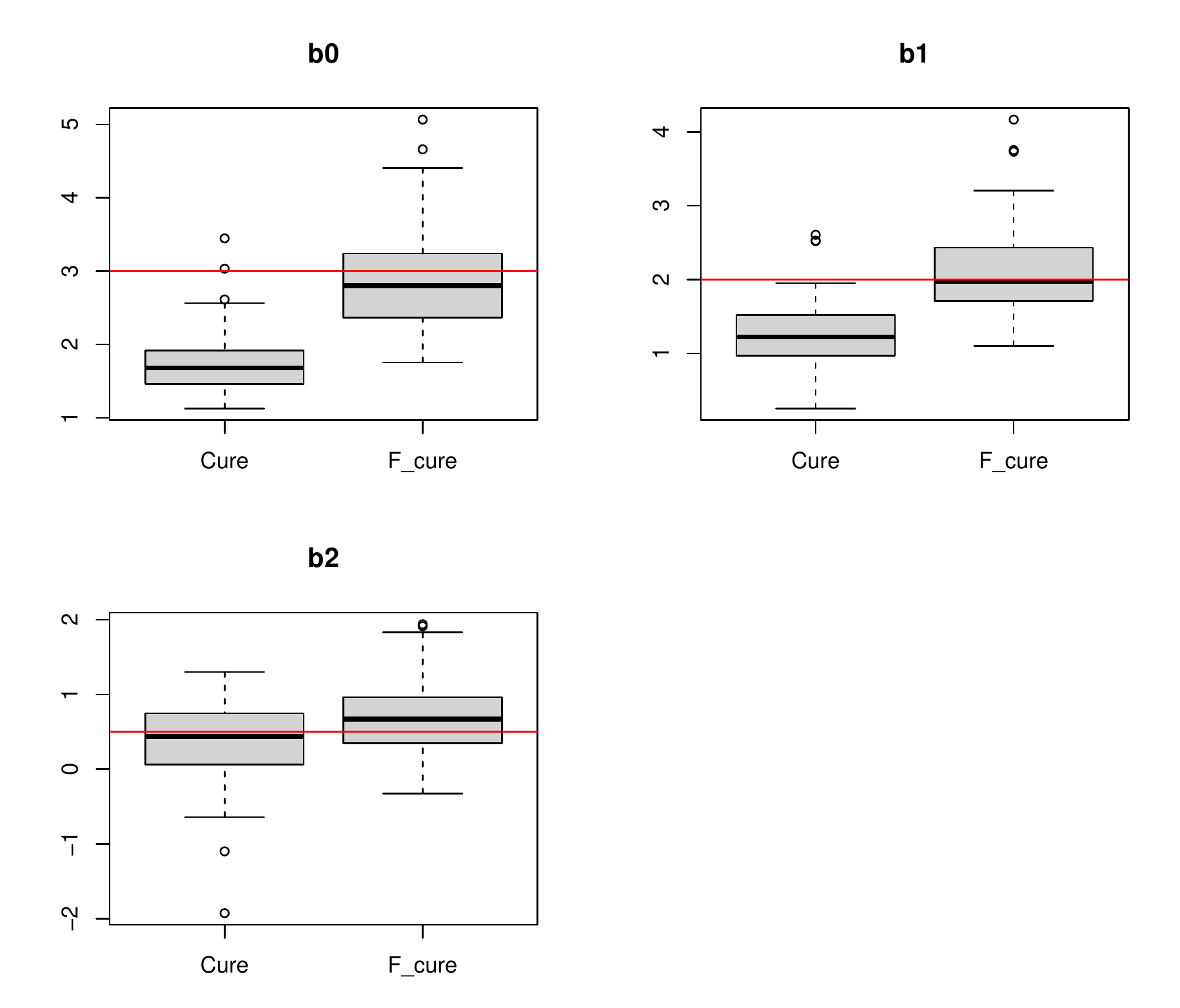}
\caption{Boxplots of $\hat{b}_0,\hat{b}_1,\hat{b}_2$ from the mixture cure model (Cure) model with scalar covariates and the proposed FPHMC method (F-cure), scenario B, n=300. The red solid line indicates the true value of the parameters.}
\label{fig:fig3b}
\end{figure}

\begin{figure}[H]
\centering
\includegraphics[width=1\linewidth , height=1\linewidth]{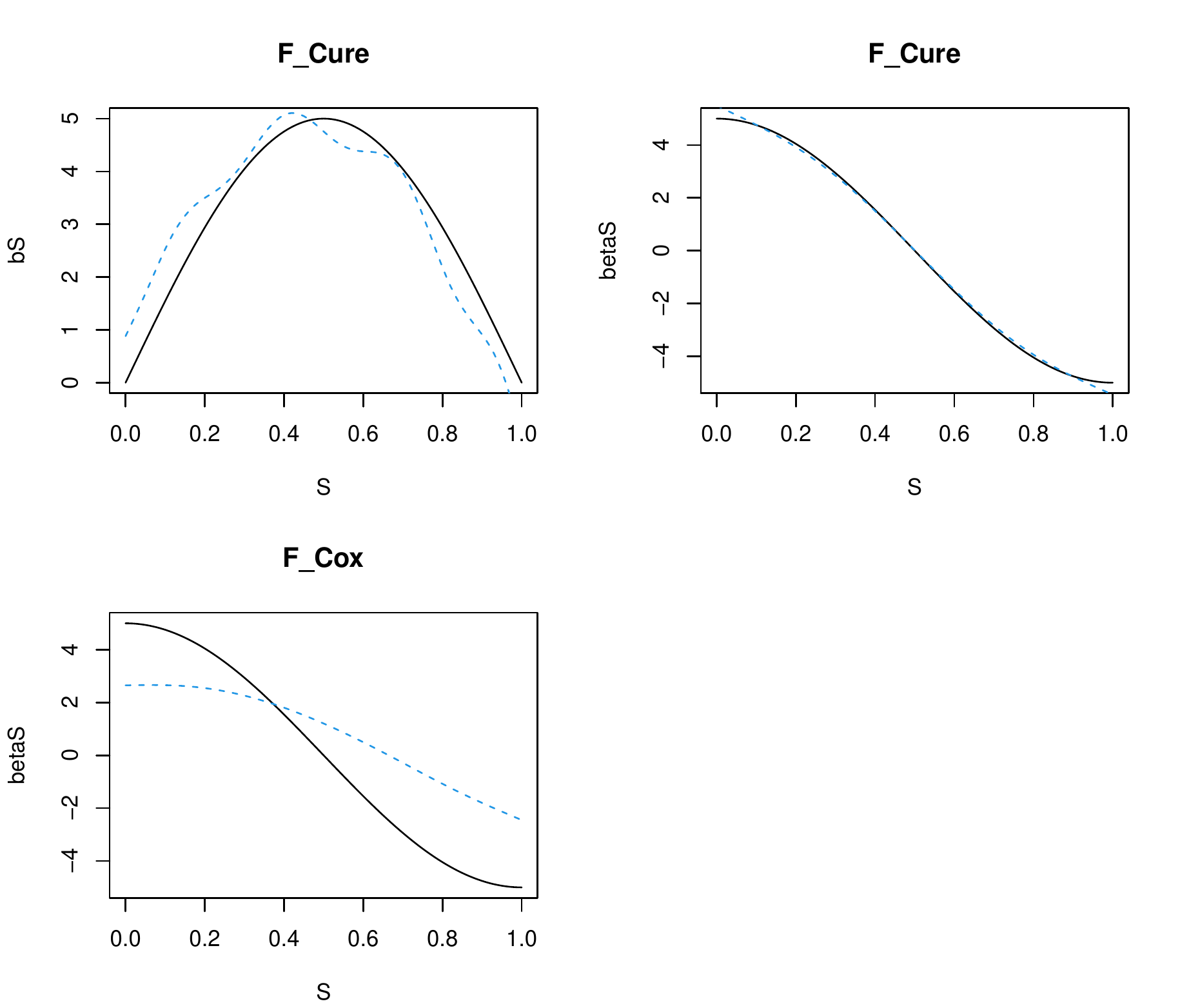}
\caption{Top left: True incidence coefficient function $b(s)$ (solid) and estimated $\hat{b}(s)$ averaged over 100 M.C replications (dotted) from the FPHMC method, Top right: True latency coefficient function $\beta(s)$ (solid) and estimated $\hat{\beta}(s)$ averaged over 100 M.C replications (dotted) from the FPHMC method, Bottom left: True latency coefficient function $\beta(s)$ (solid) and estimated $\hat{\beta}(s)$ averaged over 100 M.C replications (dotted) from the linear functional Cox method, scenario B, n=300.}
\label{fig:fig4b}
\end{figure}

\begin{figure}[H]
\centering
\includegraphics[width=1\linewidth , height=1\linewidth]{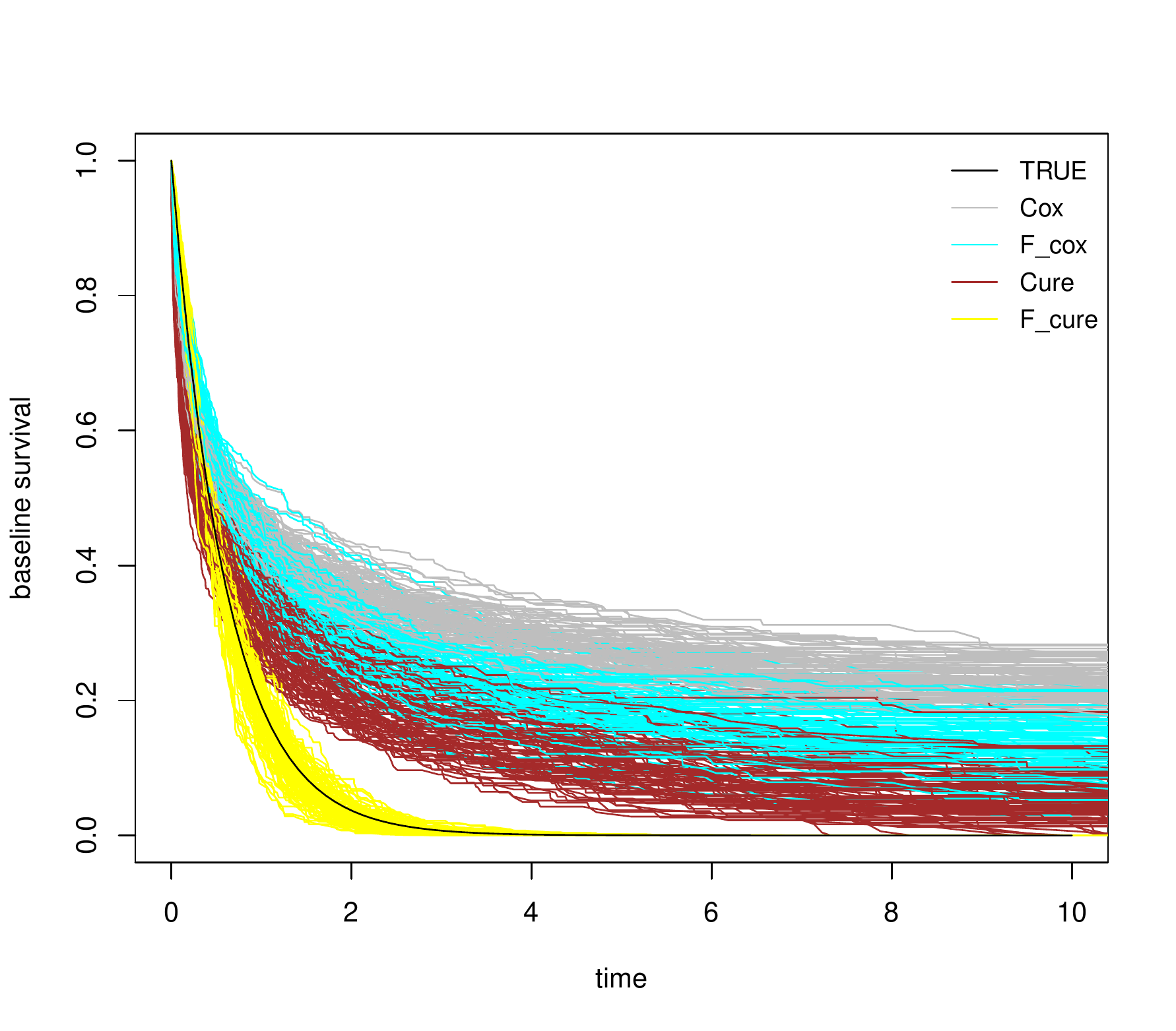}
\caption{Displayed are the true baseline survival function (black curve) and estimates of the baseline survival function $S_0(t)$ from the Cox model and mixture cure model (Cure) model with scalar covariates, linear functional cox model (F-Cox) and the proposed FPHMC method (F-cure), scenario B, n=300.}
\label{fig:fig5B}
\end{figure}

\begin{figure}[H]
\centering
\includegraphics[width=1\linewidth , height=.6\linewidth]{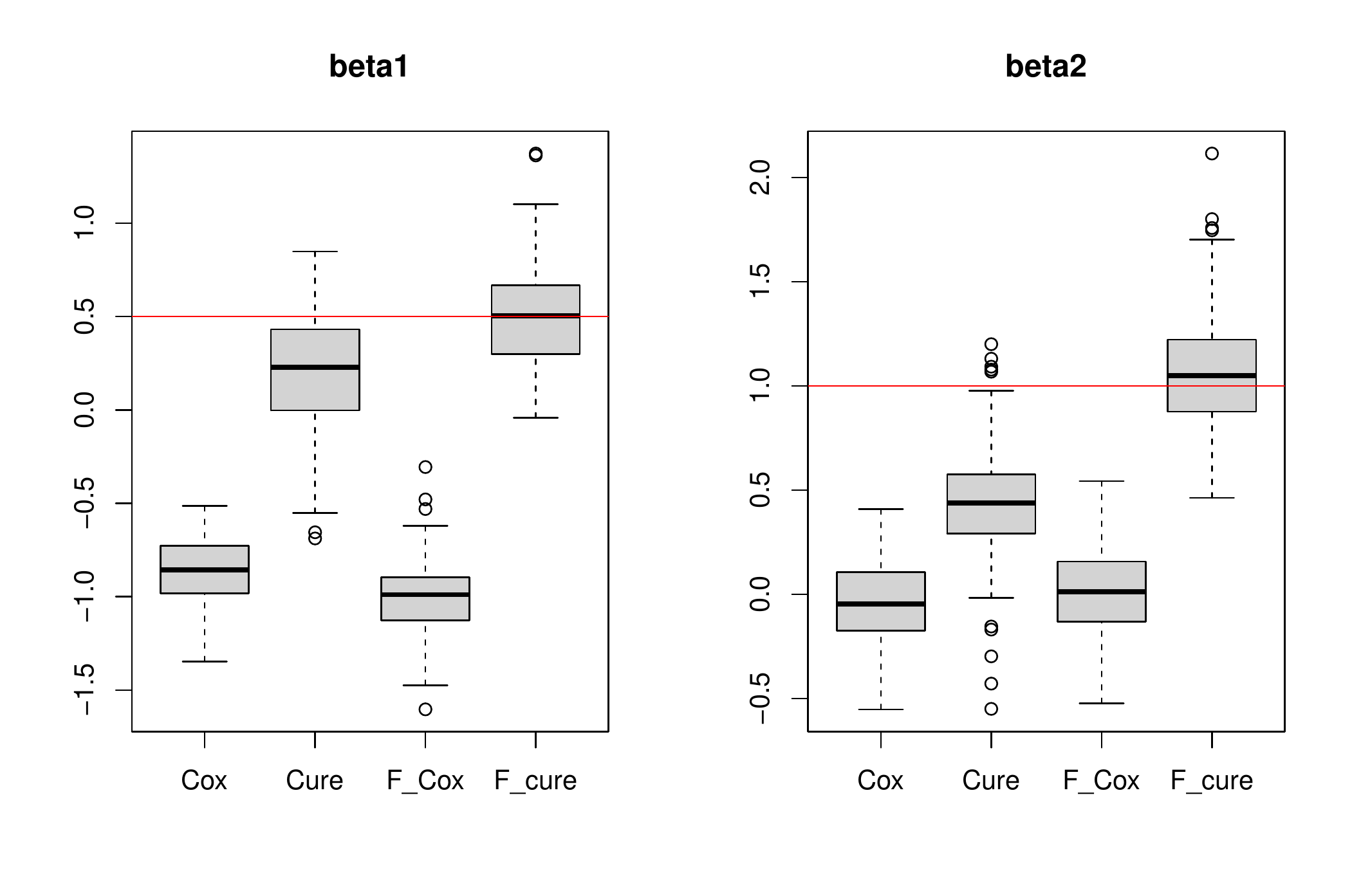}
\caption{Boxplots of $\hat{\beta}_1,\hat{\beta}_2$ from the Cox model and mixture cure model (Cure) model with scalar covariates, linear functional cox model (F-Cox) and the proposed FPHMC method (F-cure), scenario C, n=300. The red solid line indicates the true value of the parameters.}
\label{fig:fig2c}
\end{figure}

\begin{figure}[H]
\centering
\includegraphics[width=1\linewidth , height=1\linewidth]{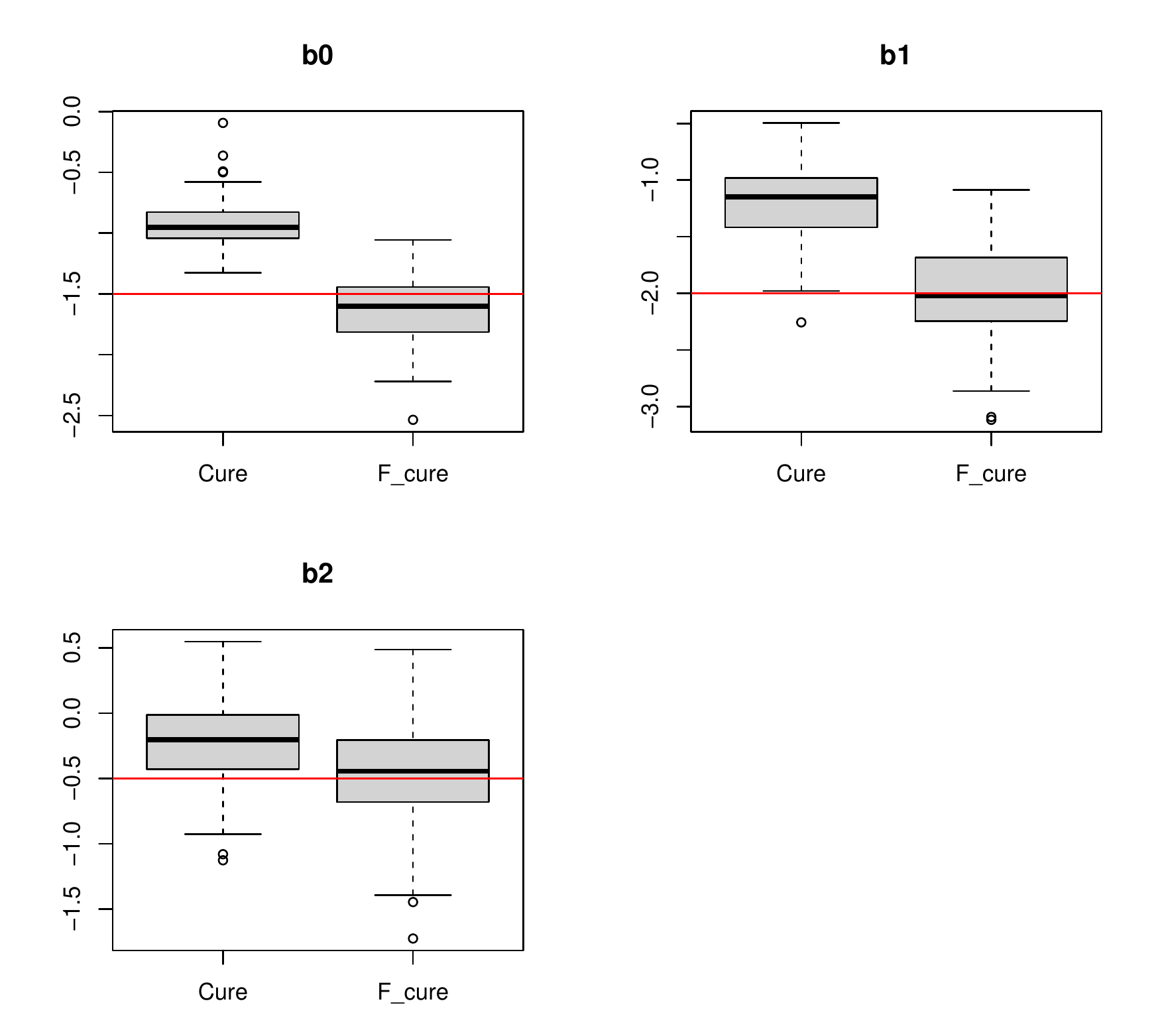}
\caption{Boxplots of $\hat{b}_0,\hat{b}_1,\hat{b}_2$ from the mixture cure model (Cure) model with scalar covariates and the proposed FPHMC method (F-cure), scenario C, n=300. The red solid line indicates the true value of the parameters.}
\label{fig:fig3c}
\end{figure}

\begin{figure}[H]
\centering
\includegraphics[width=1\linewidth , height=1\linewidth]{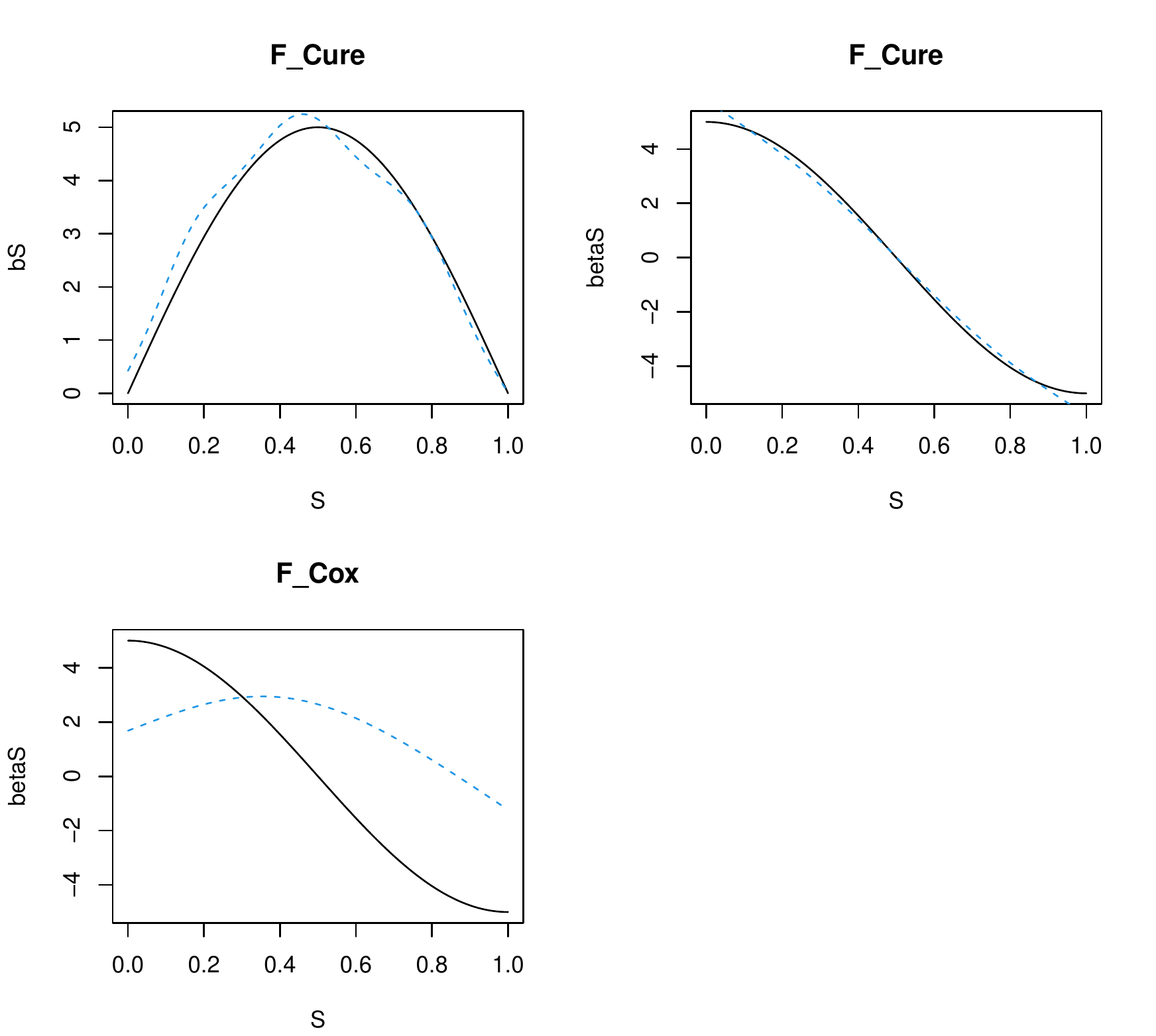}
\caption{Top left: True incidence coefficient function $b(s)$ (solid) and estimated $\hat{b}(s)$ averaged over 100 M.C replications (dotted) from the FPHMC method, Top right: True latency coefficient function $\beta(s)$ (solid) and estimated $\hat{\beta}(s)$ averaged over 100 M.C replications (dotted) from the FPHMC method, Bottom left: True latency coefficient function $\beta(s)$ (solid) and estimated $\hat{\beta}(s)$ averaged over 100 M.C replications (dotted) from the linear functional Cox method, scenario C, n=300.}
\label{fig:fig4c}
\end{figure}

\begin{figure}[H]
\centering
\includegraphics[width=1\linewidth , height=1\linewidth]{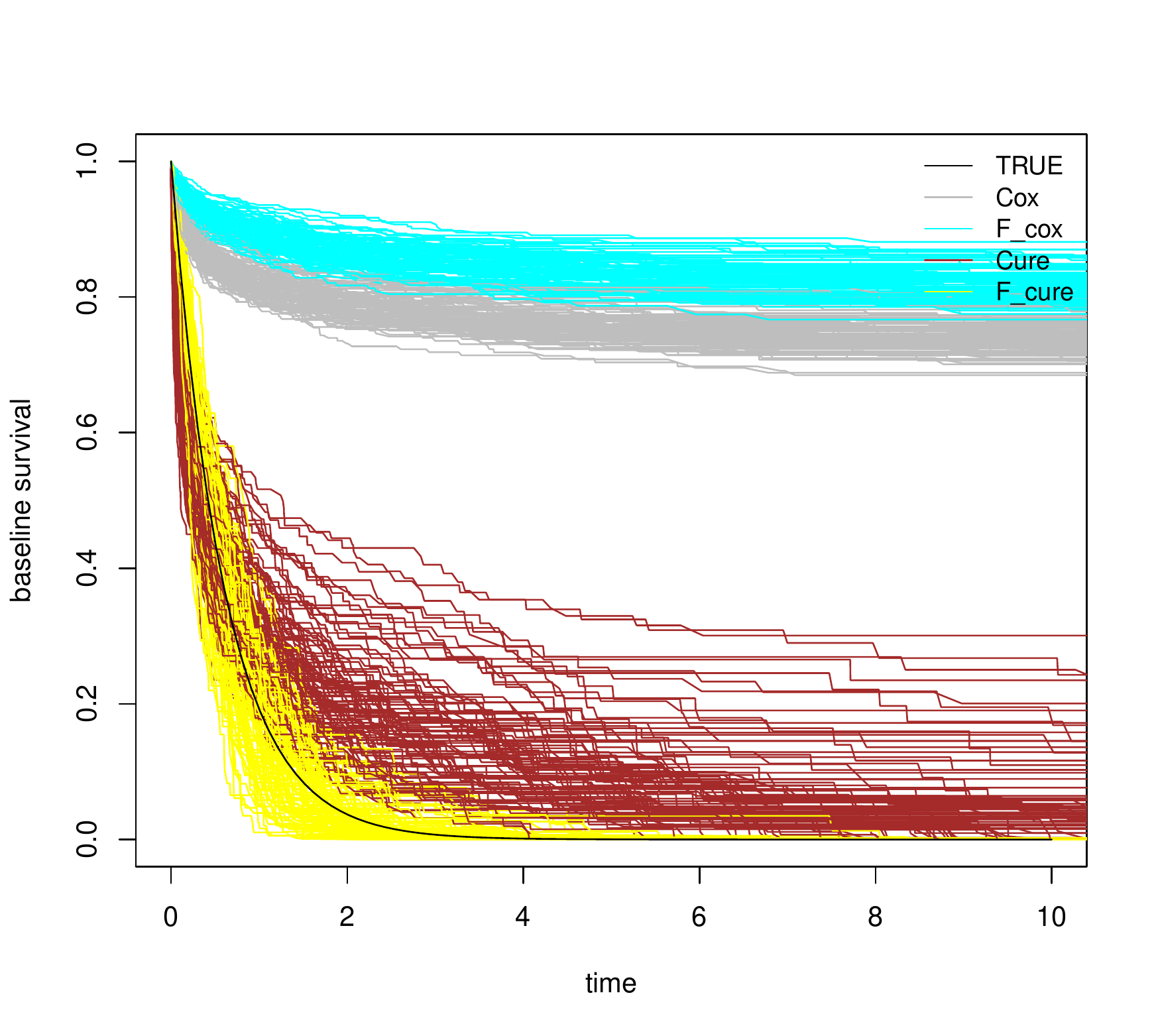}
\caption{Displayed are the true baseline survival function (black curve) and estimates of the baseline survival function $S_0(t)$ from the Cox model and mixture cure model (Cure) model with scalar covariates, linear functional cox model (F-Cox) and the proposed FPHMC method (F-cure), scenario C, n=300.}
\label{fig:fig5c}
\end{figure}

\begin{figure}[H]
	\centering\includegraphics[width=0.99\textwidth]{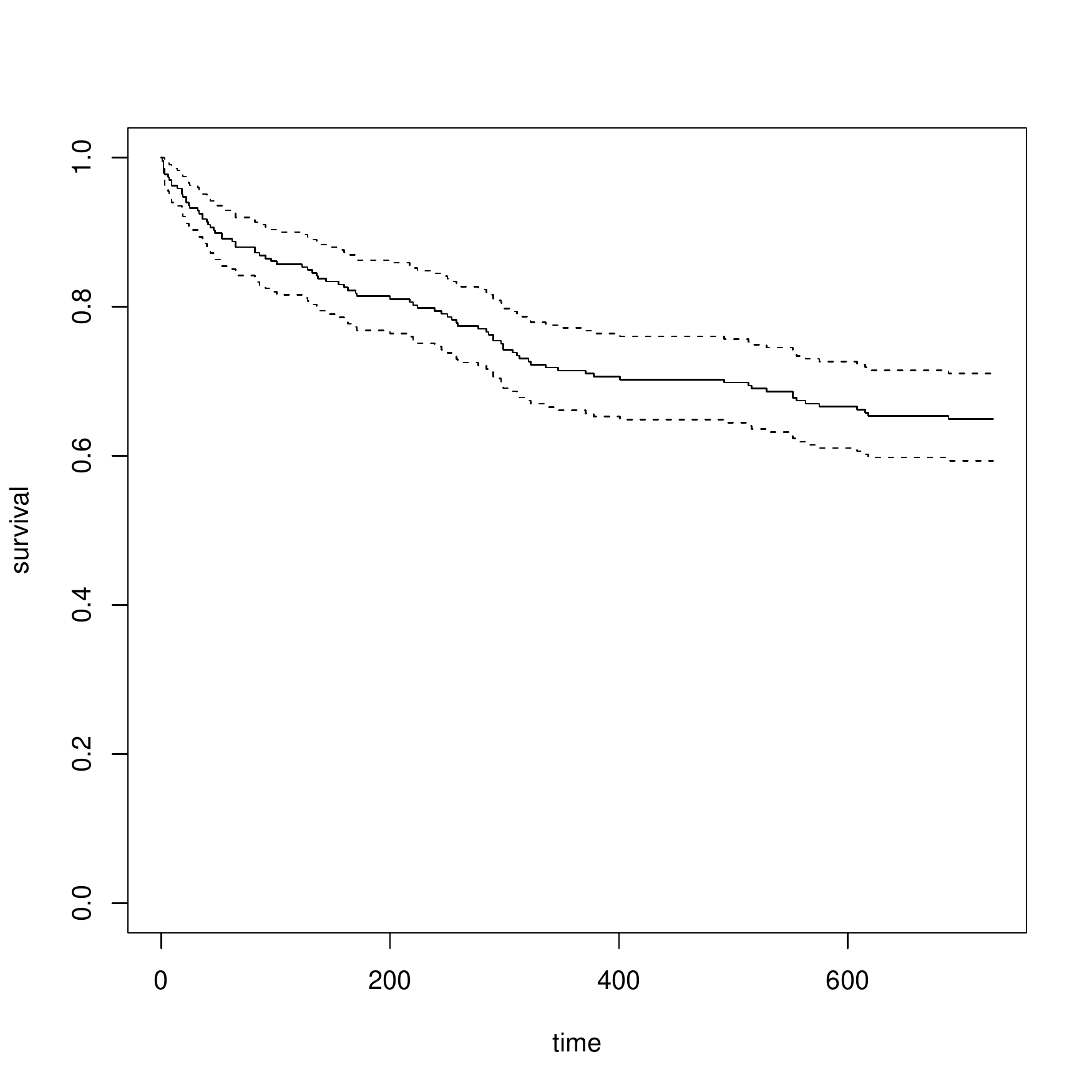}
	\caption{Marginal Kaplan Meier survival estimate in the SOFA dataset}
	\label{fig:grafcuresoga}
\end{figure}

\begin{figure}[H]
\centering
\includegraphics[width=1\linewidth , height=.7\linewidth]{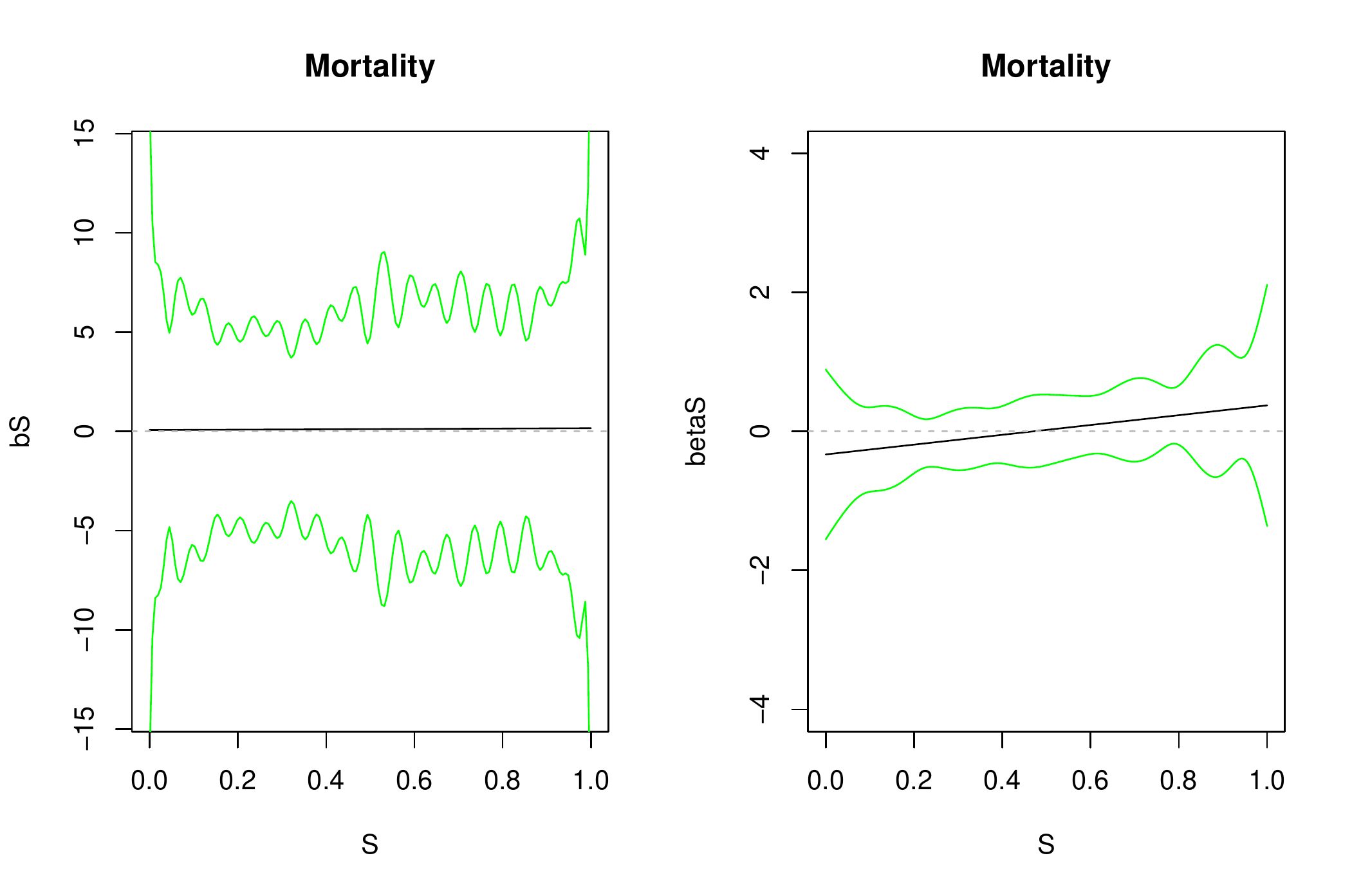}
\caption{Estimated functional effect $b(s)$ (left) and $\beta(s)$ (right) of SOFA score in the cure and latency submodel of the FPHMC model in the ICAP application along with its $95\%$ point-wise confidence interval.}
\label{fig:fig4}
\end{figure}
